\documentclass[preprint,12pt]{elsarticle}

\usepackage{amssymb}
\usepackage{amsmath}
\usepackage[]{hyperref}
\usepackage[separate-uncertainty=true]{siunitx}
\usepackage{array}
\newcolumntype{P}[1]{>{\raggedright\arraybackslash}p{#1}}
\usepackage{orcidlink}

\journal{Composites Science and Technology}

\begin{document}

\begin{frontmatter}




\title{Experimental and simulation study of resin infiltration in carbon fiber rovings\tnoteref{t1}}
\tnotetext[t1]{This document is the results of the research project (Machine Learning for Simulation Intelligence in Composite Process Design - ML4SIM) funded by the Leibniz Association.}
	

\author[1,4]{D. Burr \orcidlink{0000-0003-1351-4994} \corref{cor1}} 
\ead{dominik.burr@itwm.fraunhofer.de}

\author[2]{R. Reichenb\"acher \orcidlink{0000-0002-0852-7507} \corref{cor1}}
\ead{reichenbaecher@ipfdd.de}

\author[2,3]{C. Scheffler \orcidlink{0000-0002-6958-1304}}
\ead{scheffler@ipfdd.de}

\author[1]{K. Steiner} 
\ead{konrad.steiner@itwm.fraunhofer.de}

\author[2]{G.K. Auernhammer \orcidlink{0000-0003-1515-0143}}
\ead{auernhammer@ipfdd.de}

\cortext[cor1]{Corresponding author}

\affiliation[1]{organization={Fraunhofer-Institut für Techno- und Wirtschaftsmathematik ITWM},
            addressline={Fraunhofer-Platz 1}, 
            city={Kaiserslautern},
            postcode={67663}, 
            state={Rhineland-Palatinate},
            country={Germany}}

\affiliation[2]{organization={Leibniz-Institut für Polymerforschung Dresden e. V. IPF},
	addressline={Hohe Straße 6}, 
	city={Dresden},
	postcode={01069}, 
	state={Saxony},
	country={Germany}}           
	
\affiliation[3]{organization={Institute of Construction Materials, Dresden University of Technology},
	addressline={Georg-Schumann-Straße 7}, 
	city={Dresden},
	postcode={01187}, 
	state={Saxony},
	country={Germany}}     
       
\affiliation[4]{organization={Rheinland-Pfälzische Technische Universität Kaiserslautern-Landau},
	addressline={Gottlieb-Daimler-Straße}, 
	city={Kaiserslautern},
	postcode={67663}, 
	state={Rhineland-Palatinate},
	country={Germany}}          

\begin{abstract}
Continuous-fiber-reinforced polymers are vital for lightweight structural applications, where fiber-matrix wetting critically influences composite performance. Understanding dynamic wetting behavior during resin infiltration remains challenging, especially under realistic processing conditions. Here we investigate the dynamic wetting of commercial carbon fiber rovings by a commonly used epoxy resin through combined optical experiments and two-phase flow simulations informed by microscale roving geometries. We measure velocity-dependent advancing contact angles and observe roving geometry changes during capillary-driven resin infiltration. Simulations using microscale-derived capillary pressure and permeability parameters quantitatively reproduce the time-dependent infiltration dynamics. Our findings demonstrate that while dynamic contact angles vary with velocity, the microscale roving structure predominantly governs resin impregnation behavior. This integrated experimental and modeling approach enhances insight into fiber-resin interactions, offering a pathway to optimize composite manufacturing processes and improve material quality.

\end{abstract}

\begin{graphicalabstract}
\includegraphics[width=1\linewidth]{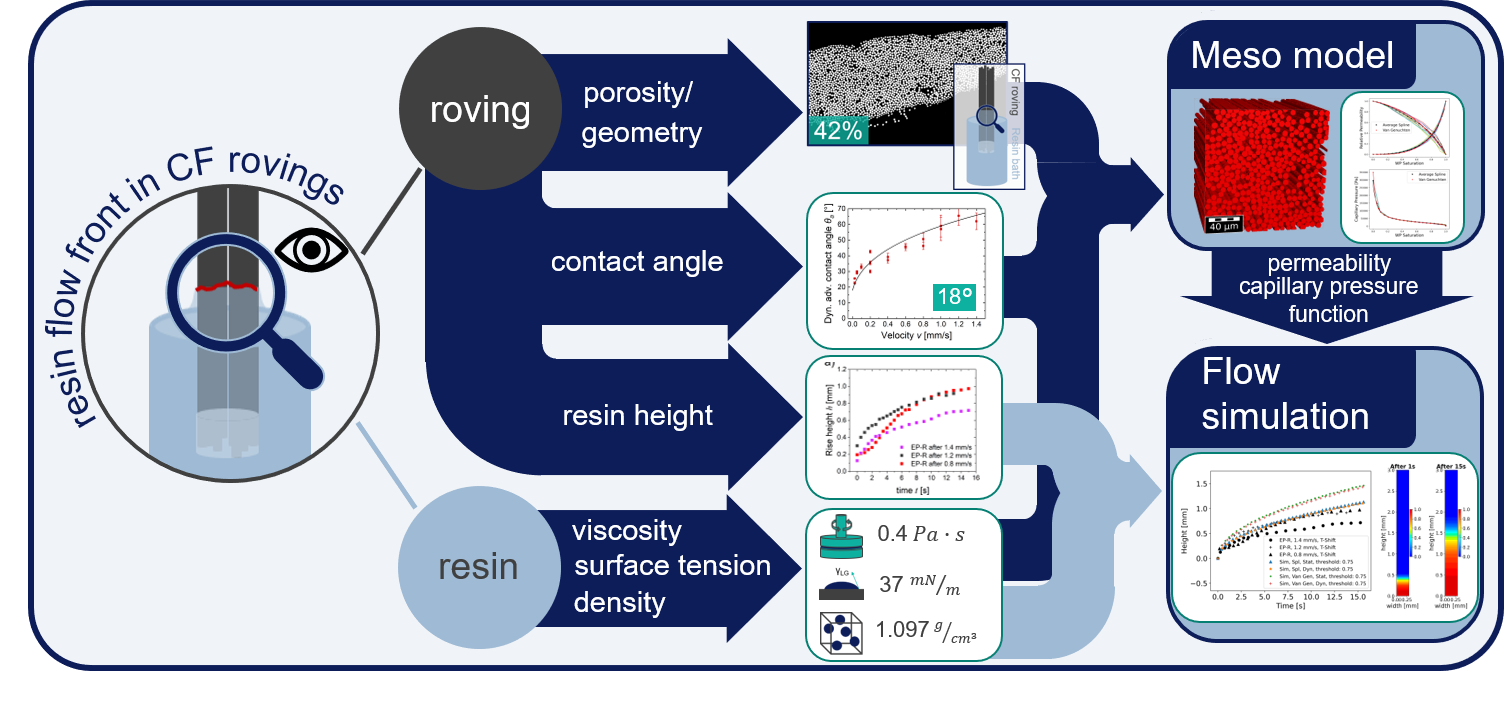}
\end{graphicalabstract}

\begin{highlights}
\item Contact angles on differently sized carbon fiber (CF) rovings vary only slightly
\item Contact angles between resin and fibers depend on the infiltration speed and are non-constant
\item The dynamic contact angle has only little influence on the considered capillary rise simulations
\item The rise height of the resin inside the CF roving is velocity dependent
\item Capillary pressure function and relative permeabilities have a significant influence on the simulation
\item Those parameters are in turn dependent on the changing roving geometry during and after wetting, the sizing chemistry and distribution and the amount and shape of the fibers influencing said geometry
\item Experimental data and simulative results could be brought to good agreement
\end{highlights}


\end{frontmatter}



\section{Introduction}
\label{sec:Intro}

\subsection{Technological motivation}
\label{subsec:Motivation}

Continuous fiber-reinforced polymers are an important group of materials for future-oriented lightweight construction applications in the automotive and transport sectors, aviation and shipping, machine elements, the construction industry and the medical sector. The combination of different fibers, usually in the form of multifilament yarns (rovings) or textiles, and different polymers (matrices) results in fiber composites with application-adapted property profiles. The main focus here is on mechanical performance with low component weight. \cite{Moskaleva2021, Maiti2022}

In an ideal composite material, the mechanical performance of the fiber-reinforced polymer is governed by the properties of the reinforcing fibers and are highly influenced by the connection between fiber and matrix via their interphase. Commonly, a sizing or coating is applied to the rovings during manufacturing, which is responsible for the chemical and physical properties of this interphase. Prerequisite for chemical bonding or physical interactions at the fiber-matrix-interphase is the physical contact, in other words, the wetting of the fibers with the liquid matrix system. A flawed wetting process during manufacturing of a composite may lead to the formation of resin-free void spaces between fibers and rovings or to simply no interpenetration of the matrix into the fiber bundle. Such defects will lead to pores in the cured composite structure that reduce the overall mechanical composite performance. \cite{Talreja2013} 

Analyzing and understanding the influence of processing parameters on the wetting behavior of a matrix system with a fiber reinforcement is expected to help gain a significant insight into composite material and processing behavior, improving composite performance and may help to facilitate simulative approaches in manufacturing and quality control. 

Since the wetting of the fiber is highly influenced by its surface chemistry, the sizing is expected to have a significant impact on wetting behavior and different surface treatments may lead to varying composite performance. \cite{Luo2011, Larson1994} Not only will the sizing influence the surface chemistry of the fibrous material but the amount and distribution of the sizing will influence the roving geometry and permeability. 

Sizings are usually applied in a range of 0.3 to 2.0$\thinspace$wt\% on rovings or textiles. \cite{OwensCorningRoving, JushiRoving, TeijinCF} The amount of sizing on the roving will determine its geometry and will change the space between the fibers that is available for the matrix to enter and permeate the roving. 

Additionally, the fiber shape and size as well as the amount of fibers in a roving will have an influence on the interaction between reinforcing fibers and matrix systems. Glass fiber manufacturers offer glass rovings with different filament diameters ranging from ca. 13 to 32$\thinspace\mu$m. \cite{DiameterGFManville, DiameterGFCPIC} Accordingly, a roving with a yarn fineness of 2400$\thinspace$tex can consist of ca. 4660 or 2070$\thinspace$filaments assuming a filament diameter of 16 or 24$\thinspace\mu$m, respectively, if a typical E-glass density of 2.58$\thinspace$g/m\textsuperscript{3} is considered. Carbon fibers usually reveal a filament diameter of around 7$\thinspace\mu$m and are offered with a fineness between 1 to 50$\thinspace$K, meaning 1.000 or 50.000 filaments in the yarn; further, the cross section of a fiber can vary from circular round to kidney-shaped. Both, the number and the shape of filaments, strongly changes the contact area between fiber and matrix and therefore influences also the wetting process during manufacturing.


During the composite manufacturing process, the thermoplastic melt or the liquid resin has to penetrate the spaces between the fibers. During infiltration processes such as resin transfer molding (RTM) or vacuum infusion processes the contact-line is moving inside the reinforcing textile structure with a certain velocity that is determined by the viscosity, surface tension and density of the matrix and processing parameters like pressure and temperature. 

With a more complex fiber shape and heterogeneous sizing distribution microscale experiments can experience a large scatter in the data and do not provide sufficient information about the macroscale fibrous material. For these reasons, studies on the mesoscale may prove more holistic in understanding fiber-matrix interactions during composite manufacturing. Gaining insights into mesoscale wetting phenomena was already explored to a certain degree in previous studies.  

In an early approach from Chwastiak \textit{et al.}, a tube coated with Polytetrafluoroethylene (PTFE) was used to hold a fiber bundle and then dipped into a wicking liquid. \cite{chwastiak_wicking_1973} However, with this approach the sample roving was deformed immensely and the geometry of it was likely changed during sample preparation. Song \textit{et al.} investigated the spreading of a drop of epoxy resin on a piece of yarn under a pre-set tension and analyzed the change in contact angle optically. \cite{song_optimizing_2023} With this approach the sample preparation can be kept non-disruptive of the inner roving geometry. However, no observation of the inner flow front of the resin could be provided. Bayramli \textit{et al.} used two experimental set-ups in order to determine the wetting by the \textit{WILHELMY} method. For axial impregnation trials, the rovings were tightly enclosed with tape and singular pieces of sealed roving were brought into contact with a wicking liquid. In a second approach a piece of roving was suspended horizontally and under tension between two brackets. The prepared piece was attached to a scale and then brought closer to a wetting liquid just until a meniscus was formed with the horizontal roving sample. The change in weight was analysed to gain insight into the resin uptake. \cite{bayramli_impregnation_1992} Especially in trials for axial impregnation the geometry of the roving was changed in order to achieve a reproducible porosity of the roving. The inner flow front of wicking liquids (silicone oils) was considered and analyzed quantitatively. Wang \textit{et al.} conducted typical dynamic single fiber contact angle measurements in combination with static optical and force based wetting experiments on rovings using water and n-hexan as wetting liquids. \cite{wang_wettability_2017} The rovings were not sealed at their end and experienced significant changes in their geometry during the experiments. A comparison of contact angles calculated from the yarn experiments and the experimental single fiber contact angles revealed substantial differences that were attributed to a change in wetting behavior due to geometrical factors when moving from the micro- to the mesoscale. 

A few papers offer insights into modeling and simulative approaches towards wetting in fibrous materials. Lukas \textit{et al.} presented a general approach towards the simulation of liquid wetting dynamics inside fiber structures. \cite{lukas_wetting_2003, lukas_computer_1997} Investigations into the wetting of a polypropylen yarn and its experimental analysis and theoretical modeling were done by Zhong \textit{et al.}. \cite{Zhong2001} Trials closer to the full scale composite manufacturing process were reviewed by Pantaloni \textit{et al.} concerning the simulation of liquid composite moulding processes with natural fibers. \cite{pantaloni_review_2020}

This work aims to lay the groundwork for a quantitative comparison of experimental wetting tests and simulations of commercial carbon fibre rovings with a commercial epoxy resin. Microscale characterisation of fibre surface roughness, sizing distribution and roving geometry informs macroscopic, two-phase Darcy flow simulations. These simulations quantitatively reproduce the time-dependent rise of resin inside the rovings, revealing that changes in porosity during infiltration and moderate influence of dynamic contact angles on flow behaviour.

\subsection{Dynamic wetting}
\label{sec:IB dynamic wetting}

The dynamics of wetting have been studied intensively over the last few decades.
For Newtonian liquids on relatively simple substrates, a generally accepted understanding of the phenomenon and its description now exists.
This has been summarised in review articles \cite{Gennes_1985aa, Marmur_2009aa, Bonn_2009aa, Snoeijer_2013aa, Butt_2015ab, Abraham_2017aa, BUTT2022101574} and textbooks \cite{Gennes:2004aa} and many others.
From a thermodynamic point of view, the contact angle of a drop on a flat, rigid and otherwise unresponsive substrate is given by the balance of surface tensions acting on the contact line, as stated by Young.\cite{Young_1805aa} 
\begin{equation}
	\cos{\theta_Y} = \frac{\sigma_{sg}-\sigma_{sl}}{\sigma_{lg}}
\end{equation}
Here, $Y$ refers to Young’s equation, where $\sigma_{sg}$, $\sigma_{sl}$, and $\sigma_{lg}$ denote the surface tensions at the solid-gas, solid-liquid, and liquid-gas interfaces, respectively.
On realistic surfaces, contact angle hysteresis is observed. Even in static conditions, this phenomenon causes the contact angle to depend on the drop's history or previous motion of its contact line. 
Between the static advancing $\theta_{adv}^s$ and receding $\theta_{rec}^s$, any intermediate contact angle can be realized based on history of the drop.
Recent reviews on contact angle hysteresis include references.\cite{Marmur_2009aa, Abraham_2017aa, BUTT2022101574} 
For moving contact lines, such as translating drops or substrates submerged in liquid, the contact line adopts a dynamic angle $\theta^d$, which varies with the velocity of contact line \cite{Ablett_1923aa}.
Similar to static cases, distinct advancing ($\theta_{adv}^d$) and receding ($\theta_{rec}^d$) angles exist for moving contact lines.

Two basic approaches have been developed to describe dynamic contact angles.
The molecular kinetic model \cite{BLAKE_1969aa} considers the jumps of the wetting liquid molecules on the substrate, deriving a dynamic contact angle as a function of contact line speed and the parameters of thermally activated molecular jumps, as well as the equilibrium contact angle.
An alternative approach starts from a hydrodynamic description of contact line motion.\cite{Moffatt_1964aa}
However, the solution to the Navier–Stokes equation obtained in this way exhibits divergent energy dissipation at the contact line.\cite{Huh:1971aa}
This divergence can be eliminated by allowing nanoscopic slip of the liquid on the solid substrate.
As demonstrated by Cox\cite{COX:1986aa} and Voinov\cite{Voinov:1976aa}, the dynamically advancing contact angle, $\theta_{adv}^d$ , can be described as follows:
\begin{equation}
	\left.\theta_{adv}^d\right.^3 = \left.\theta_{adv}^s\right.^3 + 9 \text{Ca} \ln \left(\frac{\alpha l_o}{l_i}\right) 
	\label{eq:Dyn_Contact_Angle}
\end{equation}
Where $\text{Ca} = (v \eta)/\sigma_{lg}$ is the capillary number, $\eta$ is the dynamic viscosity of the liquid, $\alpha$ is a numerical constant of order one, and $l_i$ and $l_o$ are the inner and outer length scales of the hydrodynamic solution. It is convenient to combine the logarithmic term with the friction parameter $\lambda = \ln \left[(\alpha l_o)(l_i)\right]$.\cite{Henrich:2016aa}

Equation \ref{eq:Dyn_Contact_Angle}  illustrates an important phenomenon. 
For high receding contact-line velocities, when the solid is pulled out of the liquid, the contact angle approaches zero and a wetting film remains on the solid.\cite{Landau:1942aa} 
In the inverse case, for high advancing contact-line velocities, the dynamic contact angle approaches $180^\circ$. 
And, for even higher advancing contact-line velocities, air entrainment is observed.\cite{WILKINSON:1975aa} 
These scenarios also occur in the dynamic wetting of fibers, and the maximum velocity of advancing contact lines has been demonstrated both experimentally\cite{Simpkins_2003aa} and theoretically.\cite{Shing-Chan_2011aa}

A different approach is to study the spontaneous wetting of (structured) surfaces, i.e., a surface is brought into contact and the capillary rise on this surface is observed as a function of time.\cite{Hauksbee_1710aa, Jurin_1717aa, Lucas1918, PhysRev.17.273, Quere:1997ab, FRICKE2023133895}
Initially derived for the spontaneous dynamics in capillaries, the same concepts can be applied to grooved surfaces,\cite{romero_yost_1996, Thammanna-Gurumurthy:2018aa, KUBOCHKIN2022101575, Bamorovat-Abadi_2022aa} porous materials,\cite{Marmur_1997ab, HEINZ2024125117, Khalil_2021aa} powders,\cite{Siebold_1997aa} fibers,\cite{10_1177_0040517512471742} and even liquid-liquid wetting.\cite{Rostami_2022aa}
In principle, capillary rise in corner is unbounded, if the contact angle is low enough.\cite{10_1073_pnas_63_2_292} 
This changes if the structured substrate is moved into the pool with a finite velocity. 
The capillary rise in corners shows a finite rise height.\cite{GERLACH2021126012, PhysRevFluids.7.114002}
In the case of wetting on fibers or rovings, the capillary action can also change the geometry, i.e., the distance between the fibers.\cite{Duprat_2022aa, PhysRevFluids_10_040501}

\subsection{Modeling background}

To efficiently simulate a wetting process of a composite material, we utilize a macroscopic two-phase flow model. The primary advantage of a macroscopic model is that it does not need the complete geometric information of a porous medium at the pore-scale. Instead, in these models the effect of the pore-scale geometry on the average flow is modeled by effective material parameters. These can be calculated on pore-scale geometries of small sections of the porous material, for instance small sections of a roving. The geometry for which the wetting process is to be simulated, in this work a roving, can be discretized using a relatively small number of grid cells. This is possible because there are no restrictions imposed by the size of the pores, because the pores do not appear explicitly in the macroscopic geometry. This leads to lower runtimes and less memory consumption during simulations. 

The macroscopic model used is based on the two-phase Darcy equations for Newtonian incompressible fluids. These read
\begin{flalign} \label{two_phase_darcy_comp_2fluid_satEq}
	\frac{\partial \phi S_d}{\partial t} &= - \nabla \cdot u_d, \; d \in \{ w, n \} ,  \\ \label{two_phase_darcy_comp_2fluid_velEq}
	u_d & = - \frac{k_{rd} K_0}{\mu_d} \left( \nabla p_d - \rho_d g \right), \; d \in \{ w, n \} , \\ \label{two_phase_darcy_comp_2fluid_satConst}
	S_w + S_n &= 1 ,  \\ \label{two_phase_darcy_comp_2fluid_capPress}
	p_n - p_w &= p_c,
\end{flalign}
where $d \in \{ w, n \}$, $w$ and $n$ denoting the wetting and non-wetting phase, respectively. In the equations (\ref{two_phase_darcy_comp_2fluid_satEq} - \ref{two_phase_darcy_comp_2fluid_capPress}) $u_d$ is the phase velocity, $p_d$ is the phase pressure and $S_d$ is the phase saturation. The phase saturation $S_d$ indicates what fraction of the pore space is occupied by phase $d$. Moreover, $\rho_d$ denotes the phase density, $\mu_d$ the phase viscosity and $k_{rd}$ the relative phase permeability. The variable $K_0$ is the absolute permeability, $\phi$ is the porosity and $p_c$ is the capillary pressure function of the porous material. The physical parameters $\phi$, $\rho_d$ and $\mu_d$ are determined by physical experiments. The effective material parameters $K_0$, $k_{rd}$ and $p_c$, which are saturation dependent, are calculated by efficient simulations on small pore-scale geometries as explained in section \ref{sec:Char_Mat:MicroModelling}. 

The two-phase Darcy equations were introduced as an extension of the single-phase Darcy law. Several works \cite{hornung1996homogenization, whitaker1986flow, hassanizadeh2005upscaling} derived the two-phase Darcy equations by volume averaging. For these derivations it is necessary to assume the existence of a representative elementary volume (REV) and if such a REV exists, is often not clear. An alternative that does not need the existence of a REV is the derivation by pore-scale energy dynamics that was done in McClure \textit{et al.} \cite{mcclure2022relative}.

For the simulation of capillary-driven flows the Richards equation \cite{farthing2017numerical} could be used instead of the two-phase Darcy equations. The Richards equation is a simplification of the two-phase Darcy equations, where only capillary pressure and gravity forces are considered and the influence of the air to the flow of the wetting phase is neglected. 

The capillary pressure function describes the influence of the surface tensions and contact angle to the flow on the mesoscale. The capillary pressure function as well as the relative permeabilities depend not only on the current distribution of the phases, but also on the history of the flow. The dependence on the history of the flow is known as hysteresis (see \ref{sec:IB dynamic wetting}) and because of that capillary pressure and relative permeability functions of imbibition and drainage experiments are different. As shown in McClure \textit{et al.} \cite{mcclure2018geometric} the hysteresis of the capillary pressure function can be removed by including additional state variables. \\
Moreover, the capillary pressure function is dependent on the flow dynamics. In Hassanizadeh \textit{et al.} \cite{hassanizadeh2002dynamic} these dynamical effects are incorporated by adding an additional term to the stationary capillary pressure function, so that equation \eqref{two_phase_darcy_comp_2fluid_capPress} becomes
\begin{flalign} \label{dynCapPress}
	p_n - p_w &= p_c^{stat} - \tau \frac{\partial S_w }{\partial t}  .
\end{flalign}
In this work, we only use static capillary pressure functions because determining the dynamic material coefficient, $\tau$, is complex and prone to uncertainties.
We refer to Li \textit{et al.} \cite{li2022review} for a review of the experimental methods to determine $\tau$. In this work only imbibition experiments are considered. As a consequence only imbibition relative permeabilities and capillary pressure functions are used and no hysteretic behavior needs to be modeled. Because of that we can assume that these functions are just dependent on the phase saturation. 

\section{Materials and methods}
\label{sec:Mater_Methods}

\subsection{Fiber materials}
\label{sec:Mater_Methods:Exp:fiber}
Unsized carbon fiber roving SIGRAFIL\textcopyright C T50-4.0/240-UN was procured from SGL Carbon SE. 
Three sized carbon fiber rovings, Tenax\textcopyright-J/E HTA40 E13 6K 400tex, Tenax\textcopyright-E HTS40 F13 12K 800tex and Tenax\textcopyright-J/E STS40 F13 24K 1600tex, were procured from Teijin Carbon Europe GmbH. The three different rovings have differences in the amount of fibers as well as differences in the sizings that were applied by the manufacturer. HTA40 is provided with an epoxy-compatible sizing, HTS40 with a polyurethane-compatible sizing and STS40 with a thermoplastic-compatible sizing. Addtitionally, a woven textile G-weave made from 3K carbon fiber rovings equipped with epoxy-compatible sizing was procured from Lange + Ritter GmbH. All three sized rovings contain fibers with a circular cross-section. The unsized roving as well as the textile contain fibers with a kidney-shaped cross section. An Overview of all fiber materials used is given in table \ref{tab:fiber-materials}. 

Rovings treated with different sizings were chosen in order to be able to show the influence of roving properties (such as different surface modification) on the wetting behavior during resin impregnation. In a real-life scenario the epoxy-compatible-sized roving HTA40 would be chosen for the use in combination with epoxy resin systems which is why this roving was chosen as a baseline to use for the simulative approach.

\begin{table*}
	\centering
	\begin{tabular}{|p{0.09\textwidth}|P{0.17\textwidth}|P{0.18\textwidth}||P{0.1\textwidth}|P{0.18\textwidth}|P{0.09\textwidth}|}
		\hline
		abbr. & full name & manufacturer & material type & sizing & amount of fibers \\
		\hline
		unsized  & SIGRAFIL\textcopyright-C T50-4.0/240-UN & SGL Carbon SE & roving & none & 50k \\
		EP-R  & Tenax\textcopyright-J/E HTA40 E13 6K 400tex & Teijin Carbon Europe GmbH & roving & epoxy-compatible & 6k \\
		PU-R  & Tenax\textcopyright-E HTS40 F13 12K 800tex & Teijin Carbon Europe GmbH & roving & polyurethane-compatible & 12k \\
		TP-R  & Tenax\textcopyright-J/E STS40 F13 24K 1600tex & Teijin Carbon Europe GmbH & roving & thermoplastic-compatible & 24k \\
		Textile  & G-weave & Lange+Ritter GmbH & woven textile & epoxy-compatible & 3k \\
		\hline
	\end{tabular}
	\caption{Overview of fiber materials used in this study, including manufacturer, type of sizing and amount of fibers.}
	\label{tab:fiber-materials}
\end{table*}

\subsection{Resin system}
\label{sec:Mater_Methods:Exp:Resin}
An commonly used epoxy resin system was chosen for the wetting experiments. EPIKOTE\texttrademark$\thinspace$Resin MGS\texttrademark$\thinspace$ RIMR135 and EPIKURE\texttrademark$\thinspace$ Curing Agent MGS\texttrademark$\thinspace$ RIMH137 were procured from Hexion Inc. 

This resin is a widely used epoxy system and therefor suitable to explore wettability phenomena in a system that is close to the industrial manufacturing reality. 

The two components RIMR 135 / RIMH137 were weighed with a weight ratio of 100:30 in a 12 ml disposable polypropylene container and mixed with a programmable Speedmixer (Hauschild, Speedmixer DAC 150SP). 

\subsection{Atomic force microscopy (AFM)}
\label{sec:Mater_Methods:Exp:AFM}
AFM measurements were conducted to quantify the surface roughness of the differently sized carbon fiber rovings. Multiple fibers from each roving were prepared on a microscopy slide by attaching them via double sided tape with only light pressure. Multiple spots on each fiber were measured using tapping mode to gain 3x3$\thinspace$$\mu$m topography images of the fiber surface. The measurements were done using a Dimension ICON (Bruker-Nano, USA) and a Dimension V (Bruker-Nano, USA, previously Veeco) with a cantilever Tap300-G (BudgetSensors, Bulgaria) with a nominal spring constant of 40$\thinspace$N/m, a nominal resonance frequency of 300$\thinspace$Hz and a nominal tip radius of 10$\thinspace$nm. For all measurements, a resolution of 512$\thinspace$px and a scan rate of 0.5$\thinspace$Hz were applied.\\
Using Gwyddion\cite{gwyddion} the topography data was cropped and flattened to correct the curvature of the fiber. Finally, profile lines were drawn perpendicular to the fiber axis and the roughness was averaged for all samples of one fiber type to gain the average roughness \textit{Ra}, or arithmetic average of profile height deviations from the mean line. 

\subsection{Scanning electron microscopy (SEM)}
\label{sec:Mater_Methods:Exp:SEM}
Samples for SEM were prepared by cutting sticky carbon platelets (PLANO Leit-Tabs d=12 mm) into two pieces and gluing both pieces onto a metal pen sample plate, leaving a gap between the pieces of carbon platelets. The fibers were then positioned across the gap. The samples were sputtered with Platinum under vacuum with a LEICA EM SCD 500. The SEM measurements were done using a Carl Zeiss NTS Ultra plus scanning electron microscope with the following settings:
\begin{itemize}
	\item Acceleration voltage EHT = 3$\thinspace$kV\\[-4ex]
	\item WD (Work Distance) = 7-9$\thinspace$mm\\[-4ex]
	\item Aperture Size = 30$\thinspace$$\mu$m\\[-4ex]
	\item Detector SE2\\[-4ex]
	\item Magnitude of 100x-15000x\\[-4ex]
\end{itemize}

\subsection{Tensiometer - surface tension measurements}
\label{sec:Mater_Methods:Exp:tensiometer}

We determined the surface tensions of the resin system using the pendant drop method with an OCA 35 (DataPhysics, Filderstadt, Germany). The required density of the samples for this method was determined with a density meter DMA 38 manufactured by Anton Paar (Graz, Austria). 

\subsection{Contact angle measurements}
\label{sec:Mater_Methods:CA}

In general, the two methods described for wetting characterisation can be used to measure the wetting (contact) angle.

The \textit{WILHELMY} technique is a force-based wetting method that uses a high-precision tensiometer. In this method, a single fiber or roving moves into a test liquid (resin) and the resulting weight difference is detected. \
The goniometric method involves extracting the contact angle from side-view images.

Generally speaking, for contact angles $ \theta < \SI{70}{\degree}$, the tensiometric method generates more precise results if the length of the contact line is known.
However, since carbon fibers have a typical groove structure on the surface and can have a non-circular cross-section, errors in the tensiometric method can be significant.
Additionally, the fibers would need to be extracted from the rovings. There is a risk that this method would damage the sizing interphase and leave the fibers not fully covered with sizing, resulting in heterogeneous surface properties that would influence the contact angle measurements.
To avoid these sources of uncertainty, we used the intact rovings for contact angle measurements instead.
The rationale behind this approach is that, for finite contact angles, the optically observed meniscus on the side of the roving is dominated by the wetting properties of the outermost carbon fibers of the roving.

These contact angle measurements are complemented by time-dependent capillary rise measurements of the resin inside the roving, taken using the same setup.

We prepared the roving for all contact angle and capillary rise measurements using the following procedure. First, we cut a \SI{15}{\centi \meter} piece of the roving and carefully laid it straight on a substrate. To avoid structural changes to the roving when cutting the sample to its final length, the cutting points were stabilised using Tipp-Ex Aqua (Soci\'et\'e Bic, Clichy, France). The cut ends were then sealed with Tipp-Ex Aqua. 

\begin{figure*} [h]
	\centering
	\includegraphics[width=1\linewidth]{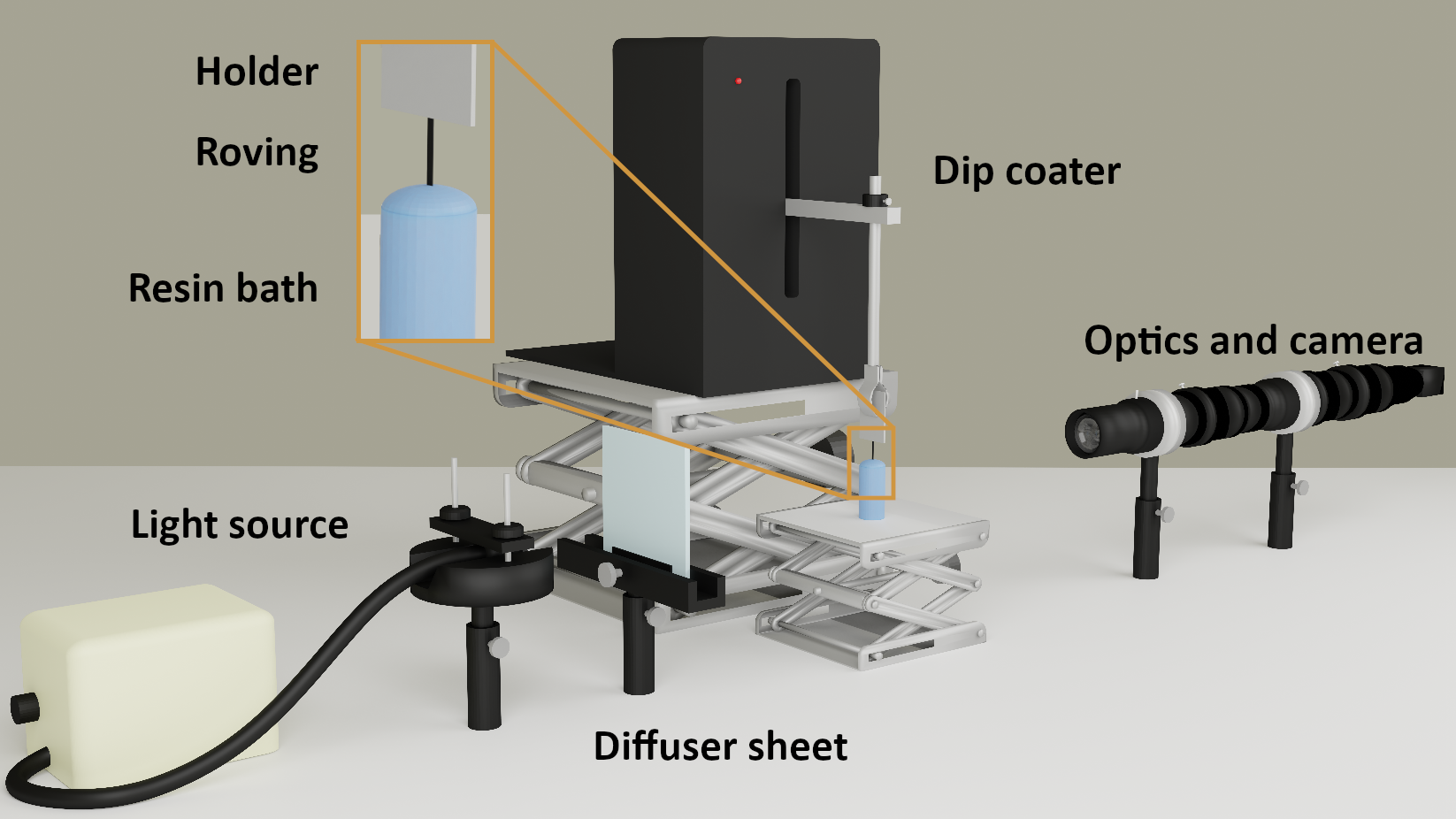}
	\caption{\label{fig:setup} The roving and the resin bath were homogeneously illuminated through a light source (KL 1500, Schott, Germany) and a diffuser sheet. We controlled the velocity of the roving with a dip coater with velocities up to \SI{1.4}{\milli \meter \per \second}.  }
\end{figure*}

Finally, the roving was carefully fixed to an aluminium holder that could be connected to the clamp of a dip coater, see Fig.~\ref{fig:setup}.

For the contact angle measurements, a cylindrical glass beaker with a volume of about \SI{10}{\milli \liter} was slightly overfilled with resin. The roving was then driven into the beaker at the desired velocity using the dipcoater, and the shape of the meniscus and the height of the capillary rise were recorded using a camera (The Imaging Source, DMK 23UP1300, Bremen, Germany) and cursom-designed objective (Thalheim-Spezial-Optik, Pulsnitz, Germany). 

The time-dependent and velocity-dependent contact angles and rise heights were extracted from the obtained videos either using custom-made software or manually using ImageJ\cite{imageJ}. If the two methods disagreed, the manual measurement was given priority.  
Following a change in velocity, the contact angle did not respond immediately, but only after a certain relaxation time. 
This relaxation can be split into a fast initial relaxation followed by a slow relaxation. We have taken the contact angle \SI{60}{\second} after the end of the fast relaxation. 
These measurements were repeated on three different samples. The error bars show the statistical standard deviation of the measured contact angles.

\subsection{Preparation of cross sections and light microscopy}
\label{sec:Mater_Methods:Exp:Cross-sections}
To determine the porosity of the roving, it was left in the resin bath overnight at the end of a dynamic wetting experiment to allow the resin to harden.  The entire sample was then completely embedded in another batch of epoxy resin in a circular mould, after which it was left to cure at room temperature. The sample was then ground down to the desired position in the roving.

Images of the cross sections were taken using a Leica DVM6A digital light microscope (Fa. Leica Microsystems GmbH) using a PLANAPO FOV 12.55 objective lens with a magnification of 1017x and coaxial lighting with an exposure time of \SI{31.5}{\milli\second}. The images were analyzed using  ImageJ.\cite{imageJ} In this process, the images were thresholded and binarized. The area fraction of the resin inside the rovings cross section was measured via the area fraction tool of ImageJ. 

\section{Characterization of the materials and static wetting}
\label{sec:Char_Mat}
\subsection{Characterization of the fiber materials}
\subsubsection{Sizing distribution and roving integrity}
\begin{figure*}
	\centering
	\includegraphics[width=1.0\linewidth]{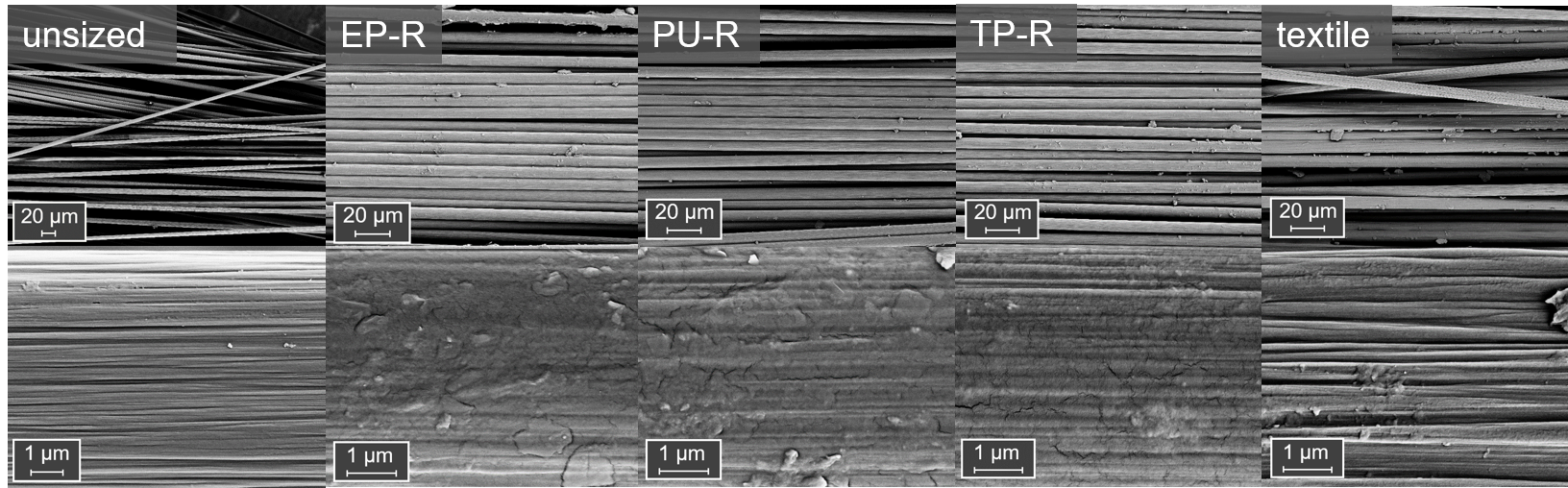}
	\caption{\label{fig::SEM} Comparison of roving geometry, roving integrity as well as single fiber surface morphology and coverage by sizing of unsized and differently sized samples including fibers from a textile.}
\end{figure*} 

The sizing chemistry as well as its distribution over the fiber surface is a pivotal factor influencing the rovings mechanical and physical properties. On carbon fibers the role of the sizing includes two major functionalities: Connecting the fibers in the roving with each other to protect the roving during processing and handling by preserving the rovings integrity and mediating the connection between reinforcing fiber and surrounding matrix in the composite. \cite{Jaber2023} The polymeric film former as the main component of the sizing is fulfilling both these functionalities while additives such as surfactants and stabilizers facilitate the processing of the sizing solution itself, partly by influencing its surface tension and wetting capabilities.

In Figure \ref{fig::SEM}, a comparison is shown between roving geometries and surface morphologies of unsized and differently sized carbon fiber specimen. Without sizing present in the roving, the roving geometry is disorganized and open with unconnected fibers running across each other leaving large empty spaces between fibers. On the surface of the fibers no coverage by sizing is visible revealing the typical surface morphology found in carbon fibers characterized by grooves running along the surface in fiber direction. Through the application of a sizing the fibers are connected along their entire axis which is made evident by a change in the roving geometry. The fibers are tightly packed and run parallel to each other with sizing polymer connecting the fibers in the interface. As a result free spaces inside the roving are minimized. The processability of the roving is improved as the fibers are held together by the sizing. Most of the fiber surface is covered with a more or less uniform layer of sizing polymer including small agglomerates. The grooves of the fiber surface are partly filled up by the sizing but the sized surface does not exhibit a perfectly smooth but rather rough appearance. Only minor differences can be seen between different sizing types. In the textile however the roving geometry is again more open and the sizing more agglomerated. This is in part due to the necessary increase in handling strain on the material as a strand of roving must be extracted from the textile weaving by hand. Differences might also be rooted in different manufacturing origins of the fibers including different sizing formulae and the additional processing step of the weaving. 

The roving geometry and its stability will have a significant influence on how the roving can interact with a wetting agent such as a resin system. A more open, unsized roving will likely diffuse and will be further disorganized by the impregnation with a resin. The resin can easily enter the free spaces between the fibers. A geometrically more stable, sized roving will stay intact during impregnation and the fibers will stay aligned during composite manufacturing. The capillaries inside the roving are stabilized by the sizing and the wetting should be more homogeneous. This will result in better composite properties after curing.  

\subsubsection{Surface roughness investigation}
\begin{figure} [h]
	\centering
	\includegraphics[width=0.75\linewidth]{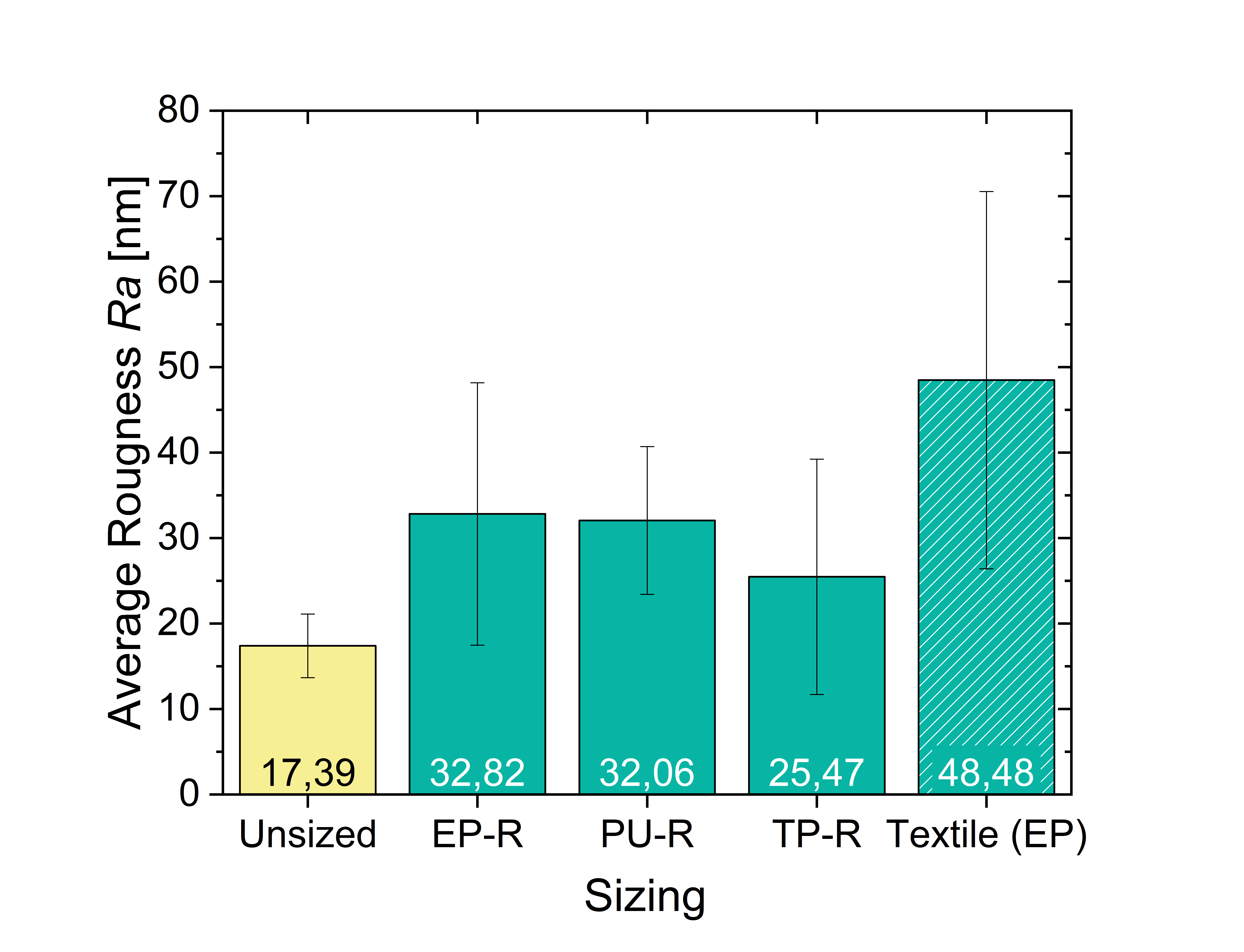}
	\caption{\label{fig::AFM} Average roughnesses Ra for unsized and differently sized fiber samples.}
\end{figure}

In figure \ref{fig::AFM} the average roughnesses of unsized and differently sized fiber surfaces are compared. Contrary to previous assumptions the roughness of the fiber surface is not necessarily lowered by the application of a sizing on carbon fibers. This partly coincides with findings in the SEM investigations where agglomerations of sizing could be seen on the fiber surfaces and a non-smooth surface was visible on all sized samples. 

Various factors can influence the surface morphology of a fiber sample including the specific manufacturer, the kind of sizing and film formers used as well as the sample preparation and method for the roughness investigation itself. The textile fibers show the highest roughness, but can only in part be compared to the rest of the samples as they stem from a different manufacturer. 
However, in all cases the application of a sizing has little influence on the surface roughness or increases it slightly. 

\subsubsection{Static advancing contact angles}

In wetting processes, roughness is only one of the physical factors that can influence the wetting behaviour of materials. The sizing chemistry and the aforementioned roving geometry are also decisive factors. To understand the wetting behaviours of fibrous materials in composites, all factors must be taken into account.

\begin{table} [h]
	\centering
	\begin{tabular}{|l||c|c|c|}
		\hline
		Sample & EP-R & PU-R & TP-R  \\
		\hline
		Contact angle $\theta_a$  & \SI{18 \pm 3}{\degree} & \SI{16 \pm 5}{\degree}  & \SI{14 \pm 2}{\degree}  \\
		\hline
	\end{tabular}
	\caption{Static advancing contact angles for all used rovings. Note that the unsized roving was not stable enough for a reliable measurement of the contact angle. }
	\label{tab:AdvStaticContacAngles}
\end{table}

We determined the static advancing contact angles as described in Sect.~\ref{sec:Mater_Methods:CA}. As the resin did not dewett from the rovings, we did not determine any receding contact angles. We repeated the contact angle measurement for three independent samples of the same roving, using the contact angle on both sides of the roving to determine the average contact angle, which is given in Table \ref{tab:AdvStaticContacAngles}. 
Comparing the static advancing contact angles with the roughness of the samples in Fig.~\ref{fig::AFM} shows a reasonable correlation between the two quantities. However, also the difference in sizing chemistry has to be considered.

\subsubsection{Changes in porosity in the roving during wetting}

When the roving was wetted with the resin, a fluid-structure interaction driven by capillary forces was generated between the resin and the fibers of the roving. 
The videos of the dynamic wetting experiments reveal that the roving with an initially flat ribbon-like shape developed an almost circular cross section in the region in which capillary rise inside the roving was observed. 
To quantify this fluid-structure interaction, we determined the porosity of the roving in three different positions.

\begin{figure*} [h]
	\centering
	\includegraphics[width=0.8\linewidth]{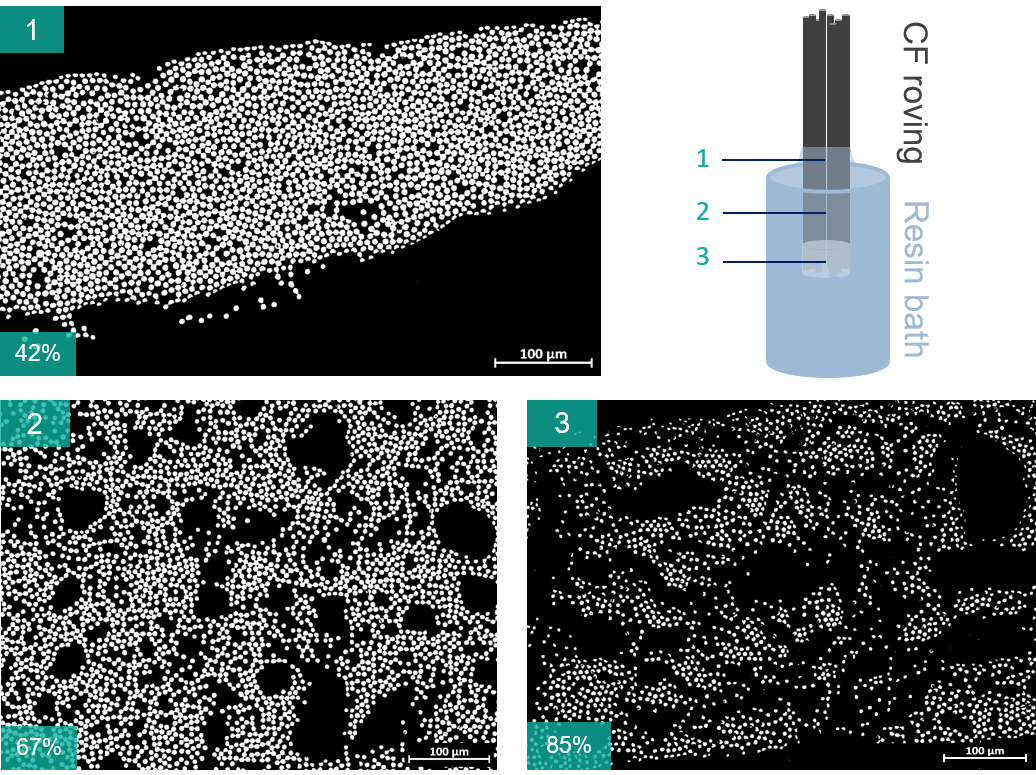}
	\caption{\label{fig::porosity} Cross sectional images of carbon fiber roving EP-R at different positions (see figure \ref{fig:cross-sections}), analyzed for area fraction of the resin (porosity of the roving). }
\end{figure*}

Cross sections of resin-wetted rovings were taken at three distinct positions as indicated in Fig~\ref{fig:cross-sections}:
1) In the sealed end of the roving.
2) The section of the roving that was completely immersed into the resin bath. 
3) Above the pool level, in the region where most of the roving deformation was observed. 
These cross sections show a strong change in inner structure of the roving, Fig.~\ref{fig:cross-sections}. This fluid-structure interaction is expected to influence the capillary rise in the roving. 
In the sealed end of the roving it is suspected that the initial dry geometry of the roving is mostly preserved. A porosity or area fraction not occupied by fibers in the cross section of the roving of 85$\thinspace$\% was observed, see Fig. \ref{fig::porosity}-3. When the roving is immersed into the resin bath, the resin is pulled up through capillary forces between the fibers changing the rovings geometry. The fibers in the roving are pulled together by the resin flow front reducing the porosity of the roving to a minimum of 42$\thinspace$\%, see Fig. \ref{fig::porosity}-1. After being fully immersed into the resin bath, below the fluid surface, the geometry of the roving is relaxed due to reduction of capillary forces. The rovings porosity increased again to 67$\thinspace$\%, see Fig. \ref{fig::porosity}-2. 

This data shows that changes in the geometry of a roving occur upon wetting with a resin system. This likely will have an influence on the impregnation behavior of the roving during composite manufacturing and in consequence on the final composite properties. The roving-resin interaction is influenced by the sizing and resin chemistry and the resins rheology and surface tension. 
According to these findings, for micromodelling a porosity of a porosity of 42$\thinspace$\% is used as that geometry represents the state in the flow front that is supposed to be simulated. 

\subsection{Viscosity and surface tension of the resin system}
\label{sec:Char_Mat:Viscosity}

\begin{figure} [h]
	\centering
	\includegraphics[width=0.75\linewidth]{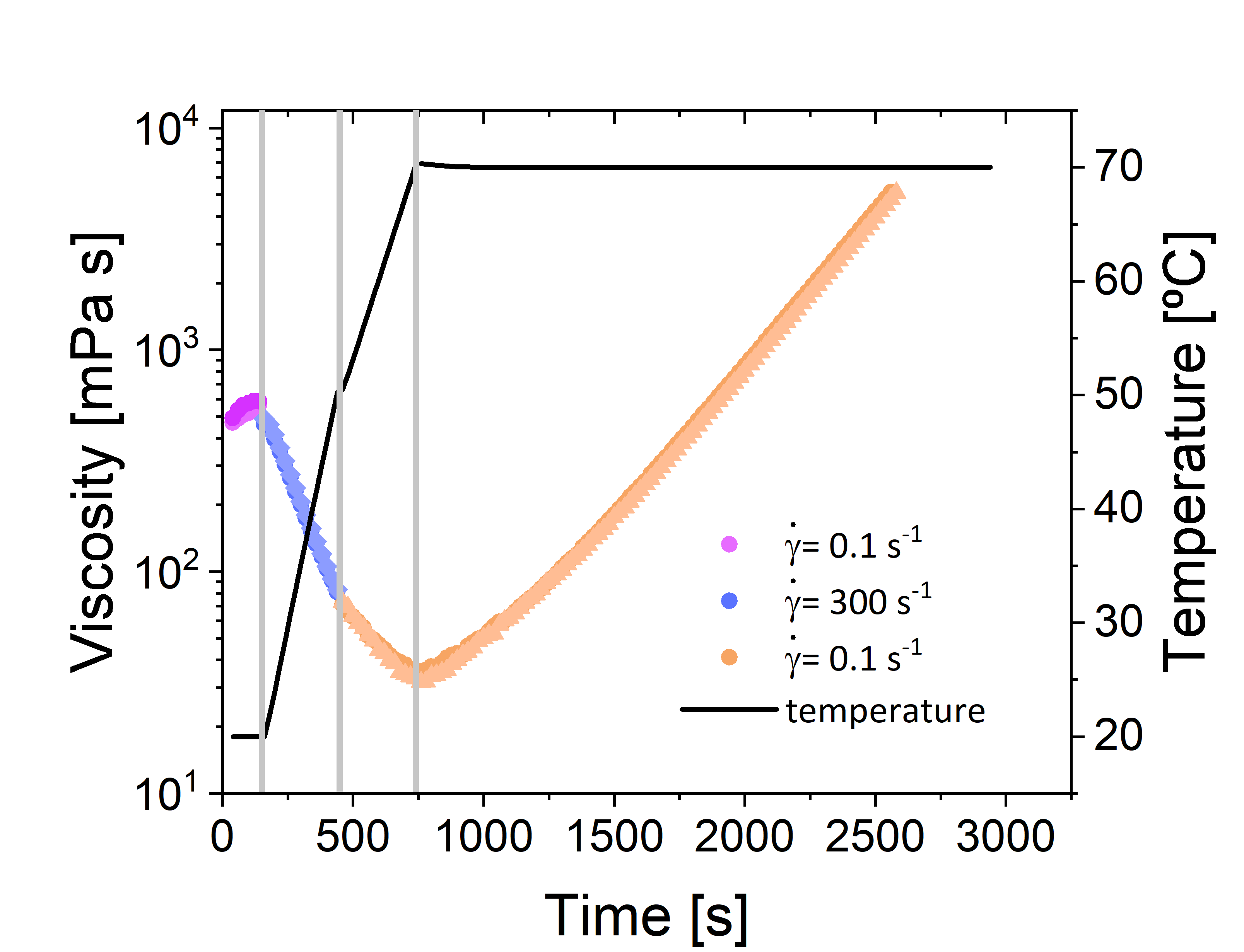}
	\caption{\label{fig:viscosity} Viscosity vs. time of a 100:30 mixture of RIMR135 and RIMH137 in a temperature profile that was inspired by a typical application situation. The two overlapping curves indicate the reproducibility of two independent measurements.}
\end{figure}

We determined the rheological properties in temperature profiles that is inspired by a classical mold-filling process. This temperature profile is split in four parts as indicated in Fig.~\ref{fig:viscosity}. In initial phase at room temperature gives the reference viscosity that is relevant for the static and dynamic wetting experiments presented in this work (0.4$\thinspace$Pas).
At elevated temperatures, the crosslinking changed the viscosity after roughly \SI{800}{\second}. All experiments at room temperature were performed in the time frame where the viscosity was not significantly changed. 

The surface tensions and densities in Table \ref{tab:SurfaceTensionDensity} were determined at room temperature for all components as described in Sect.~\ref{sec:Mater_Methods:Exp:tensiometer}. 

\begin{table*}
	\centering
	\begin{tabular}{|P{0.25\textwidth}||P{0.2\textwidth}|P{0.2\textwidth}|P{0.2\textwidth}|}
		\hline
		Sample & resin & hardener & resin:hardener (100:30)  \\[1ex]
		\hline
		Surface tension $\sigma_{lg} [$\SI{}{\milli \newton \per \meter}$] $ & $\SI{44.4 \pm 0.2}{}$ & $\SI{32.2 \pm 0.2}{}$  & $\SI{37.0 \pm 0.2}{}$  \\[1ex]
		Density $\rho [$\SI{}{\gram \per \cubic \centi \meter}$] $ & $\SI{1.143\pm 0.001}{}$ & $\SI{0.950 \pm 0.001}{}$  & $\SI{1.097 \pm 0.001}{}$  \\[1ex]
		\hline
	\end{tabular}
	\caption{Surface tension and densities of the resin and its components.}
	\label{tab:SurfaceTensionDensity}
\end{table*}

\subsection{Capillary pressure and permeability by mesoscale modeling}
\label{sec:Char_Mat:MicroModelling}

To set up simulations of the capillary rise experiments the effective material parameters are needed. These effective material parameters are the absolute permeability $K_0$, the relative permeabilities $k_{rd}$ and the capillary pressure function $p_c$. There are multiple procedures to determine these parameters by physical experiments \cite{peters2015revisiting, zhuang2017analysis}. In this work we determine these material parameters by simulations. For these simulations we use the GeoDict software from Math2Market. 

As a first step mesoscale geometries are created as voxel geometries. These mesoscale geometries have a predefined porosity. The exact arrangement of the fibers in a mesoscale geometry is random. Because of that the material parameters calculated in two different mesoscale geometries with the same porosity are still different. 

To calculate the absolute permeability $K_0$ a saturated flow through the geometry is simulated by solving the Stokes equation. After that, the mean velocity and the pressure drop in flow direction are inserted into the single-phase Darcy equation to calculate the permeability. 

\begin{figure} [h]
	\centering
	\includegraphics[width=0.5\linewidth]{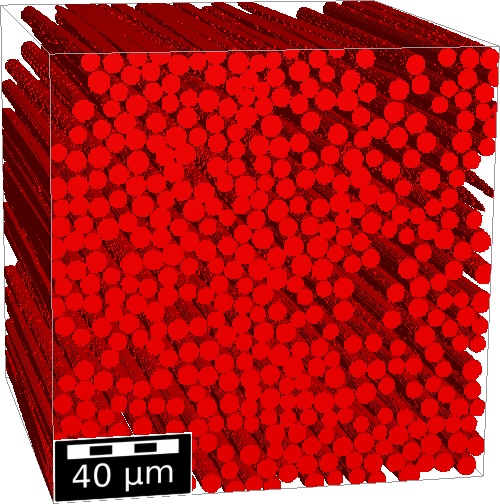}
	\caption{\label{fig::microGeo} Depicted is one of the mesoscale geometries used to calculate the capillary pressure function and permeabilities of the roving.}
\end{figure}

For the calculation of the capillary pressure function and relative permeabilities the pore-morphology method from Hilpert \textit{et al.} \cite{hilpert2001pore}, together with the generalization to not totally wetting materials from Schulz \textit{et al.} \cite{schulz2007modeling}, is used. This pore-morphology method is an efficient algorithm to generate equilibrium phase distributions in pore-scale geometries for different capillary pressure values. \\
These phase distributions can be used to calculate the stationary capillary pressure function. Therefore, only the saturation of the wetting phase of the generated phase distributions needs to be calculated. \\
Moreover, the generated phase distributions can be used to calculate the relative permeabilities. For that the Stokes equation is used to calculate a stationary saturated flow of one of the fluid phases. After that, as for the absolute permeability, the single-phase Darcy equation is used to calculate the permeability. This calculated permeability is divided by the absolute permeability to get the relative permeability of the respective phase at the phase saturation value of the stationary state. 

With these procedures, we calculate the effective material parameters of the microscale geometries of the roving. We generated ten different geometries that have a size $160 \mu m$ in all three dimensions. These geometries contain circular filaments with a diameter of about $7 \mu m$ that have all nearly the same direction. According to the experimental results, the porosity of the geometries is $\phi = 0.42$, Fig.~\ref{fig::porosity}. One of the microscale geometries is depicted in Figure \ref{fig::microGeo}. During these calculations, we use \SI{37}{\milli \newton \per \meter} as surface tension, Tab.~\ref{tab:SurfaceTensionDensity}, and $18^{\circ}$ as stationary contact angle as it was measured for the EP-R material, Tab.~\ref{tab:AdvStaticContacAngles}.

\begin{table} [h]
	\centering
	\begin{tabular}{|c|c|}
		\hline
		Mesoscale Geometry & Absolute Permeability  \\
		\hline
		$ 1 $ & $9.33 \cdot 10^{-13} m^2$ \\
		$ 2 $ & $8.42 \cdot 10^{-13} m^2$ \\
		$ 3 $ & $7.63 \cdot 10^{-13} m^2$ \\
		$ 4 $ & $8.63 \cdot 10^{-13} m^2$ \\
		$ 5 $ & $9.12 \cdot 10^{-13} m^2$ \\
		$ 6 $ & $7.57 \cdot 10^{-13} m^2$ \\
		$ 7 $ & $8.41 \cdot 10^{-13} m^2$ \\  
		$ 8 $ & $7.22 \cdot 10^{-13} m^2$ \\
		$ 9 $ & $7.15 \cdot 10^{-13} m^2$ \\
		$ 10 $ & $7.18 \cdot 10^{-13} m^2$\\
		\hline
		\hline
		Arithmetic Average & $8.06 \cdot 10^{-13} m^2$ \\
		\hline
	\end{tabular}
	\caption{ \label{tab:microMatParam_absPermeab} This table lists the absolute permeabilities calculated on ten mesoscale geometries and the arithmetic average of these ten values.  } 
\end{table}



\begin{figure} [h]
	\centering
	\includegraphics[width=0.75\linewidth]{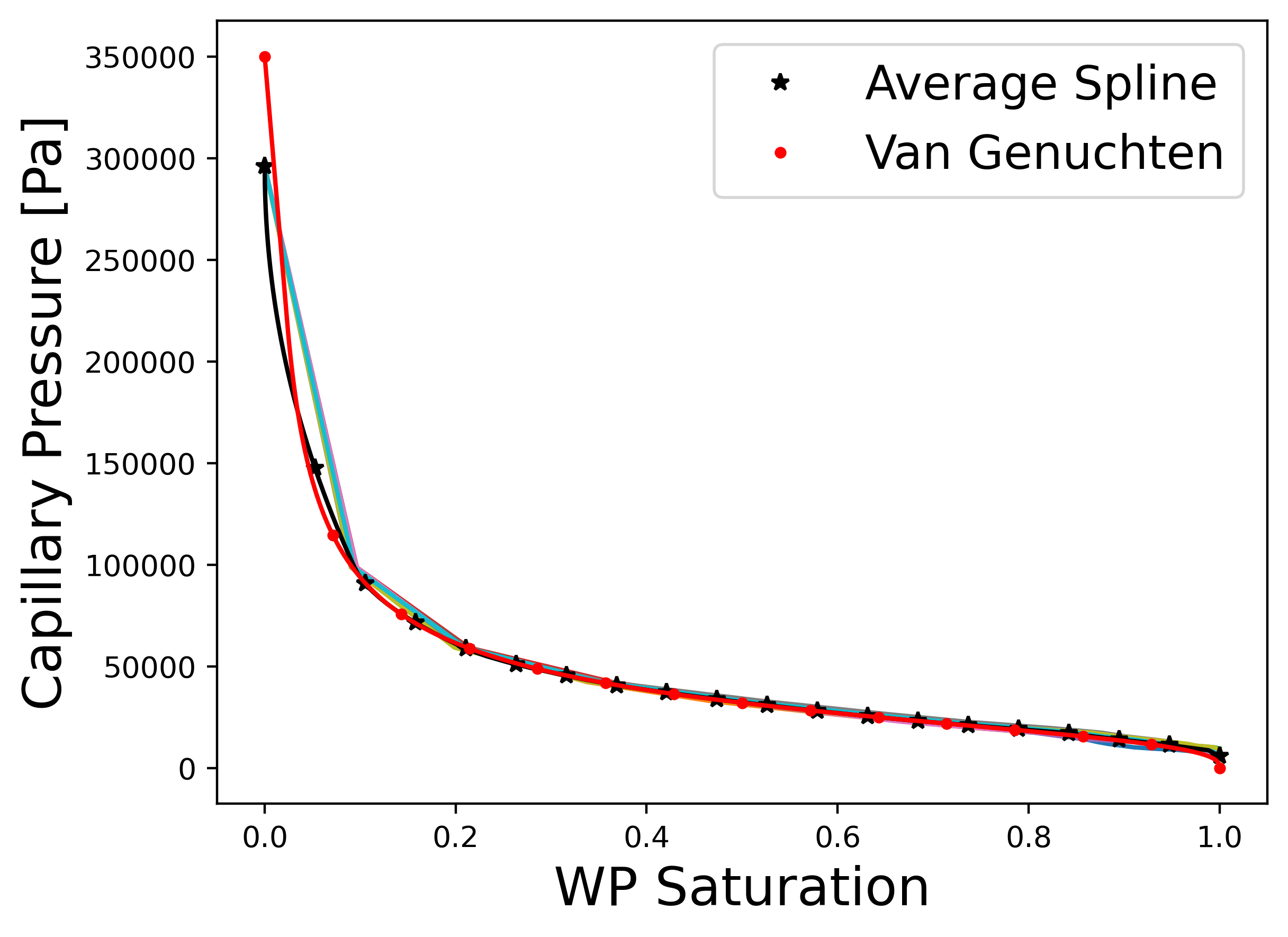}
	\caption{\label{fig:microMatParam_capPress} The image show the capillary pressure functions calculated on ten mesoscale geometries. The black line marks the average of all mesoscale results and the red line shows the Van Genuchten capillary pressure function fitted to the average function. The other lines show the results of the mesoscale geometries. The parameters of the Van Genuchten capillary pressure function are $p_e = 24875 Pa$ and $n_C = 2.72$. }
\end{figure}

\begin{figure} [h]
	\centering
	\includegraphics[width=0.75\linewidth]{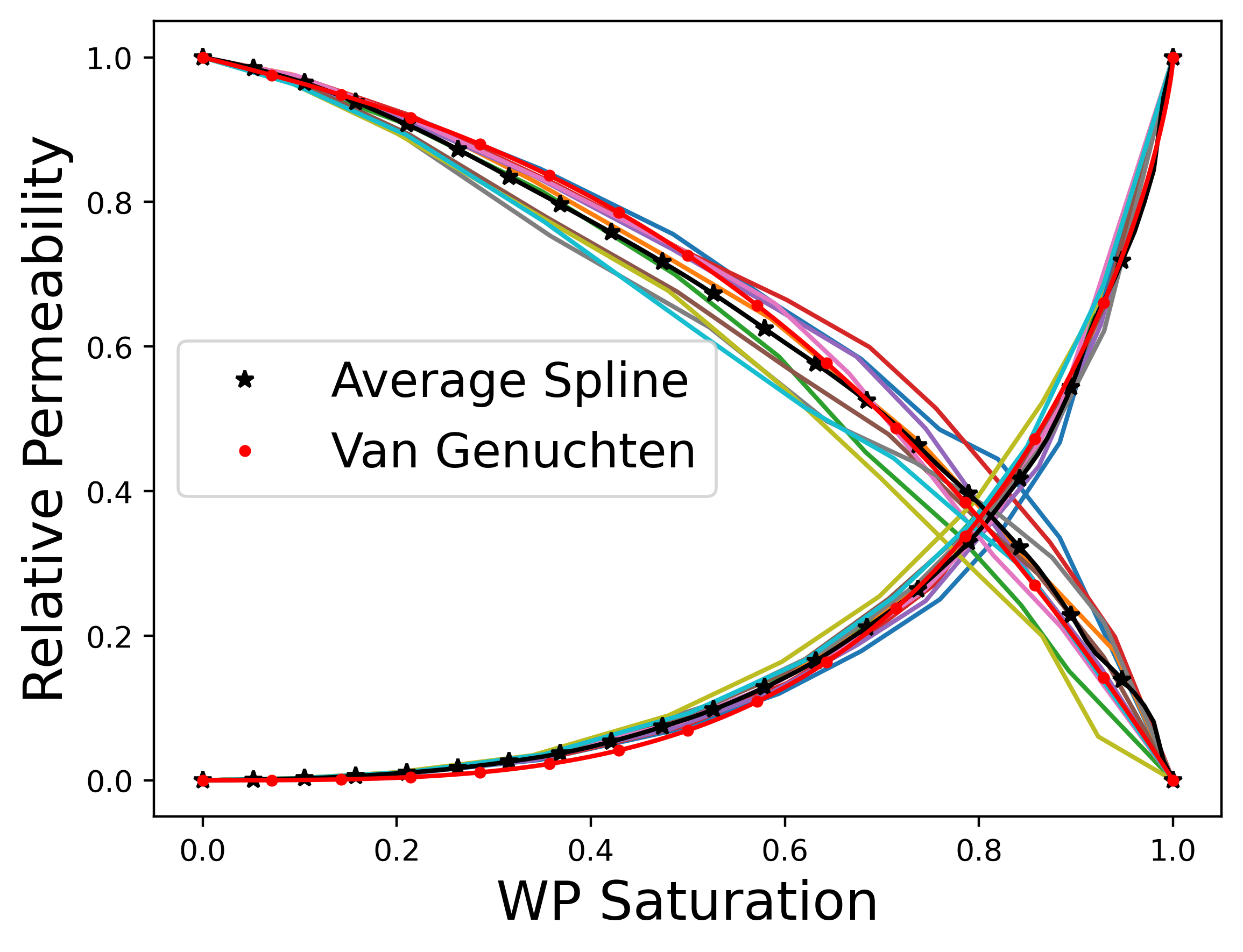}
	\caption{\label{fig:microMatParam_relPermeab} The image show the relative permeabilities calculated on ten mesoscale geometries. The black lines mark the averages of all mesoscale results and the red lines show the Van Genuchten relative permeability models fitted to the average functions. The other lines show the results of the mesoscale geometries. The parameters of the Van Genuchten relative permeabilities are $n_w = 3.94$ and $n_{nw} = 1.5$.}
\end{figure}

We calculate the averages of the effective material parameters from all ten microscale geometries. The averages are calculated as explained by Becker \textit{et al.} \cite{becker2024efficient} in Appendix A. The calculated absolute permeabilities are listed in Table \ref{tab:microMatParam_absPermeab}, the capillary pressure functions are presented in Figure \ref{fig:microMatParam_capPress}, and the relative permebilities are presented in Figure \ref{fig:microMatParam_relPermeab}.

The calculated average capillary pressure function and relative permeabilities can be directly used in the simulations as a monotone spline function of the data points. Often, a model function for the capillary pressure functions is fitted to the data points. In this work we do both variants. As model function that is fitted to the data points, we use the Van Genuchten capillary pressure function \cite{van1980closed} and relative permeabilities \cite{bear2012introduction, helmig1997multiphase}. To define these, we first need to introduce the effective wetting phase saturation, \begin{flalign}
	\bar{S}_w = \frac{S_w - S_{wr}}{1 - S_{wr} - S_{nr}} .
\end{flalign}
In the above equation $S_{wr}$ is the residual wetting phase saturation and $S_{nr}$ is the residual non-wetting phase saturation. 

The Van Genuchten capillary pressure function is given by
\begin{flalign}
	p_c (S_w) = p_e \left( {\bar{S}_w}^{- \frac{n_c}{n_c - 1}} - 1 \right)^{\frac{1}{n_c}}.
\end{flalign}
In this formula $p_e$ and $n_c$ are the parameters that need to be fitted. 

The Van Genuchten relative permeabilities are
\begin{flalign}
	k_{rw} (S_w) &= \bar{S}_w^{\frac{1}{2}} \left( 1 - \left( 1 - {\bar{S}_w}^{\frac{n_w}{n_w-1}} \right)^{\frac{n_w-1}{n_w}} \right)^2, \\
	k_{rn} (S_w) &= \left( 1 - \bar{S}_w \right)^{\frac{1}{3}} \left( 1 - {\bar{S}_w}^{\frac{n_{nw}}{n_{nw}-1}} \right)^{\frac{2 \left( n_{nw} -1 \right)}{n_{nw}}}.
\end{flalign}
In these formulas $n_w$ and $n_{nw}$ are the parameter that we fit to the calculated relative permeabilities. 

In Figure \ref{fig:microMatParam_capPress} the Van Genuchten capillary pressure function is compared to the average and individual microscale results, and in Figure \ref{fig:microMatParam_relPermeab} the same is presented for the Van Genuchten relative permeabilities.

\section{Dynamic wetting}
\subsection{Dynamic contact angles}

\begin{figure} [h]
	\centering
	\includegraphics[width=0.75\linewidth]{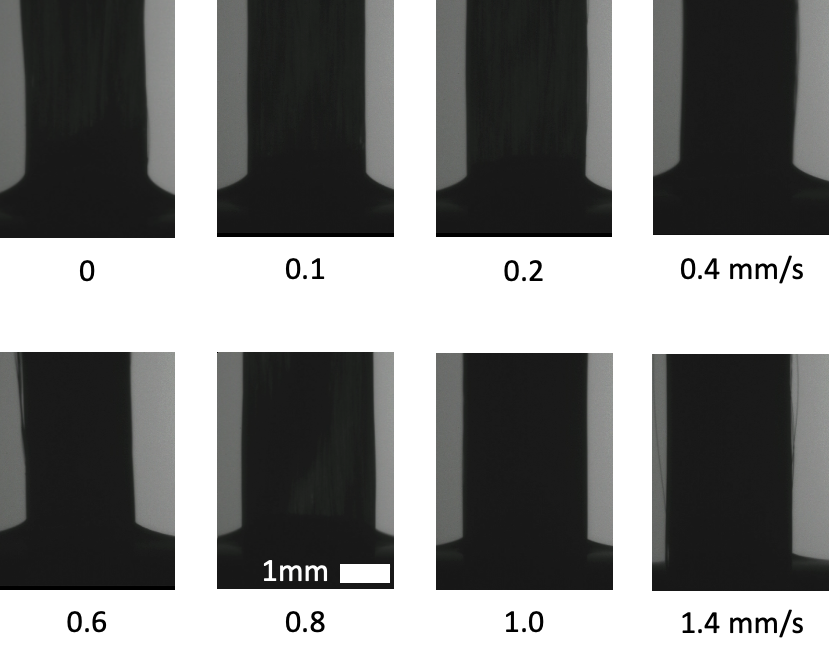}
	\caption{\label{fig:snapshots} This series of snapshots shows that as velocity ($v$) increases, so does the dynamic advancing contact angle $\theta_a(v)$. The scale bar applies to all images. 
		\textbf{GA: Check the the kind of roving taken.}}
\end{figure}

When performing dynamic contact angle measurements, i.e. measuring the advancing contact angle $\theta_a(v)$ while the roving is driven into the resin bath at a constant velocity, the advancing contact angle increases monotonously with velocity. This is illustrated in the series of images in Fig.~\ref{fig:snapshots}. This observation agrees with the expectations found in Eq.~\ref{eq:Dyn_Contact_Angle}.

\begin{figure} [h]
	\centering
	\includegraphics[width=0.75\linewidth]{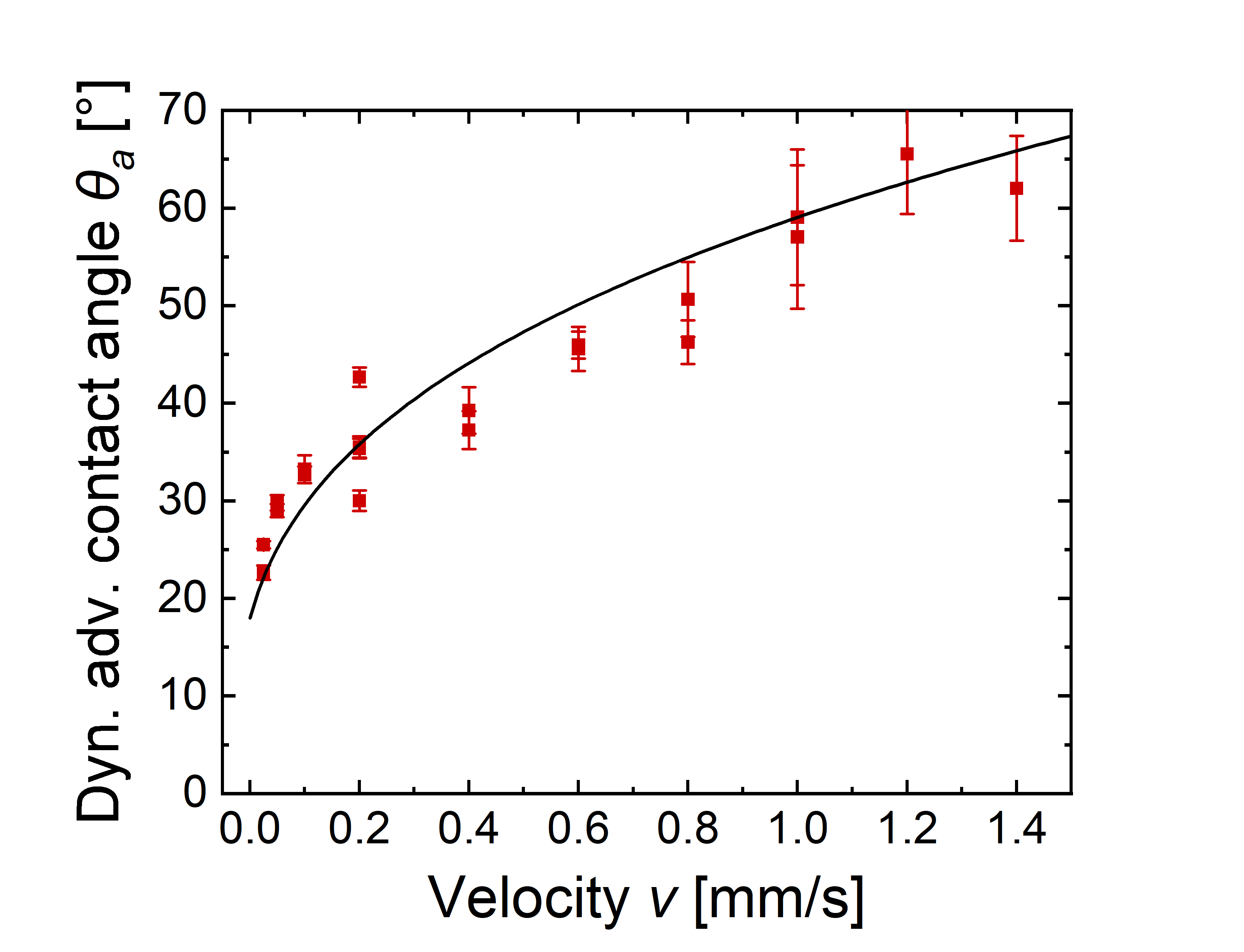}
	\caption{\label{fig:dyn_CA} Dynamic contact angles. The data scattering illustrates the heterogeneity of the rovings. The solid line is a guide to the eye using Eq.~\ref{eq:Dyn_Contact_Angle}.}
\end{figure}

The quantitative comparison of the measured dynamic advancing contact angles also matches the functional form of Eq.~\ref{eq:Dyn_Contact_Angle} roughly, see Fig.~\ref{fig:dyn_CA}. However, a couple of the assumptions underlying this equation were not met in our experiments due to the fact that wetting on a roving was considered here, rather than wetting on a flat substrate as assumed in Eq.~\ref{eq:Dyn_Contact_Angle}. This results in two major differences:
i) The roughness of the wetted surface, Fig.~\ref{fig::SEM}, induced pinning of the contact line.
ii) There was capillary rise inside the roving that advanced in front of the wetting front on the outside of the capillary. Consequently, the flow close to the contact line moving over the outside of the roving is much more complex than assumed in Eq.~\ref{eq:Dyn_Contact_Angle}.

Due to this additional effect, a direct comparison with standard modeling is not possible. 
The solid line in Fig.~\ref{fig:dyn_CA} should not be considered as a fit of data to Eq.~\ref{eq:Dyn_Contact_Angle}, but a guide to the eye. 
However, there are a couple of comparable cases. For example, in the case of drop spreading on porous substrates,\cite{Jearn01012010, ROSENHOLM20158} the dynamic contact angle also followed the $\theta^3$ dependency of Eq.~\ref{eq:Dyn_Contact_Angle}. In general, it is expected that the drag force of flow over rough surfaces filled with liquid will be reduced.\cite{Schonecker_2015aa}

\subsection{Capillary rise}

Additional to the forced wetting at the outside of the roving, spontaneous wetting, i.e. capillary rise, so called "wicking", inside the roving was observed. For this we used an additional coaxial illumination to be more sensitive to changes in the reflectance of the roving to the liquid inside the roving. We quantified this spontaneous process for non-moving rovings and for rovings moving with a constant velocity into the resin bath. 

\begin{figure}
	\centering
	\includegraphics[width=1\linewidth]{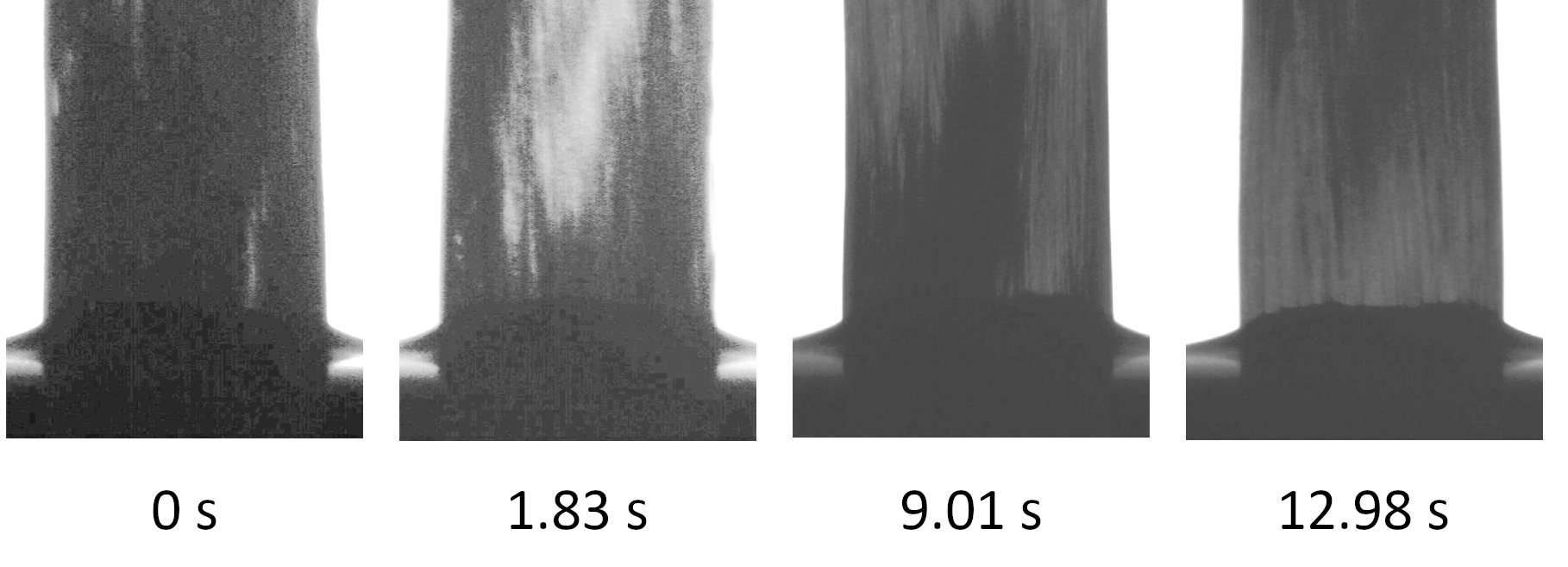}
	\caption{\label{fig:CapillaryRiseImages} Snapshots of the rise in the roving after stopping from \SI{0.8}{\milli\meter\per\second} in a EP-R roving. The corresponding plot is shown in Fig.~\ref{fig:CapillaryRise}. }
\end{figure}

\begin{figure}
	\centering
	\includegraphics[width=1\linewidth]{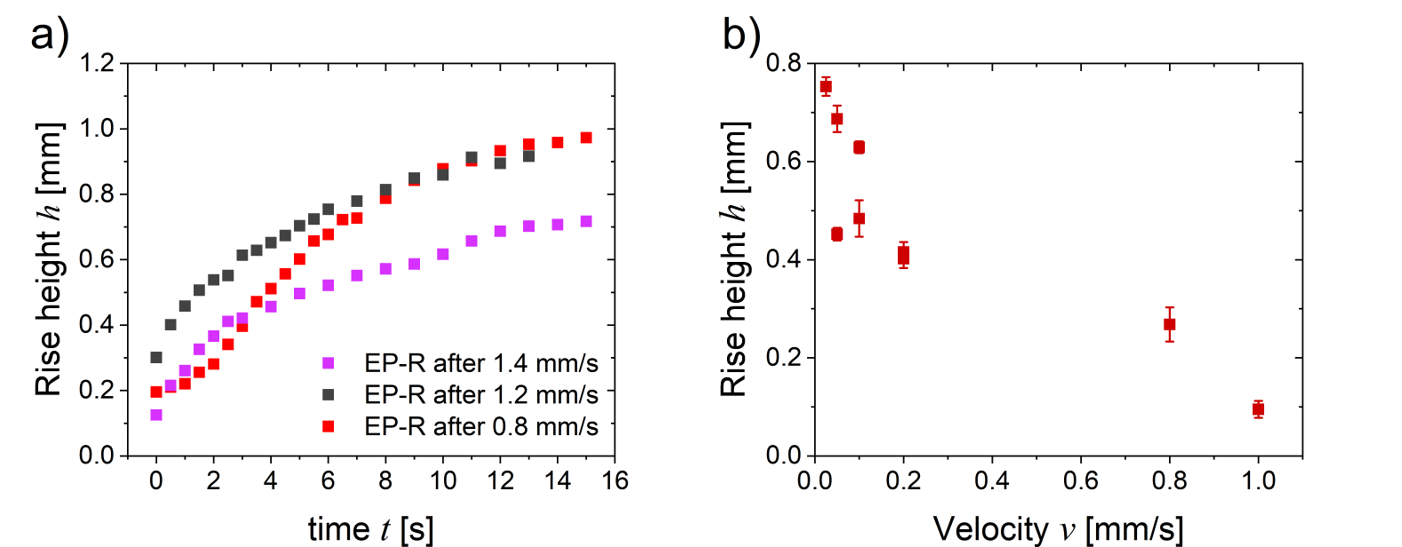}
	\caption{\label{fig:CapillaryRise} a) The rise height inside the roving for three different initial velocities before stopping the motion. b) The stationary-state rise heights at different velocities of the rovings. For all experiments EP-R was used. Data point from experiments with not sufficient optical contrast are omitted. }
\end{figure}

For non-moving rovings, we initially drove the roving with relatively high velocity, as indicated in Fig.~\ref{fig:CapillaryRise}, into the resin bath, then stopped the motion and observed the rise height above in side the roving the meniscus level the sides of the roving. 
In most of the cases, the wetting front inside the roving was not horizontal, but increased from the sides of the roving towards its center, with a central region of the roving with a (roughly) constant rise height, Fig.~\ref{fig:CapillaryRiseImages}. 
The time dependence of this rise heights (Fig.~\ref{fig:CapillaryRise}) shows two important features: 
i) The rise height starts at non-zero values for slow initial velocity of the roving. This originates from the balance between the capillary rise in inside the roving (analogue the the capillary rise in powders \cite{Siebold_1997aa, KIRDPONPATTARA2013169} and the continuous driving of the roving into the resin pool, i.e., opposite to the direction of the rise. 
ii) The rise height increases with time similar to the capillary rise as described by Lucas\cite{Lucas1918} and Washburn.\cite{PhysRev.17.273} 

In the steady state of a roving moving into the resin bath with a constant velocity, the rise height is a decreasing function of the roving velocity, Fig.~\ref{fig:CapillaryRise}b and decreases with increasing roving velocity. In the present setup, it was not possible to observe the rise height or contact angle for dynamic contact angle above \SI{90}{\degree}. However, it is safe to assume that the behavior would be similar as for single fibers driven into a liquid.\cite{Simpkins_2003aa} This implies that air entrainment into the roving is expected if the forced wetting velocity is too high.

\section{Simulative replication of capillary rise experiments}
For the simulations, the two-phase Darcy Eqs.~(\ref{two_phase_darcy_comp_2fluid_satEq} - \ref{two_phase_darcy_comp_2fluid_capPress}) are solved using an Implicit Pressure Explicit Saturation (IMPES) method with a finite volume discretization on a voxel grid. The used two-phase Darcy solver is introduced and validated in Burr \textit{et al.} \cite{burr2025timestep}

To simulate the capillary rise experiment a 2-dimensional geometry with a height of $3 mm$ an inlet at the bottom and an outlet at the top is used. To simulate the capillary rise the same pressure value is prescribed at in- and outlet, i.e., the pressure drop from in- to outlet is zero. The porous material inside the geometry is assumed to be homogeneous and is characterized by the microscale material parameters calculated in Section \ref{sec:Char_Mat:MicroModelling}. During the simulations we choose \SI{0.4}{\pascal\second} as the viscosity of the resin, Figure \ref{fig:viscosity}. As density of the resin, we use the value listed in Table \ref{tab:SurfaceTensionDensity}.

There are four simulations conducted. The first two use the capillary pressure and relative permeability curves derived directly from the monotone spline functions depicted in Figure \ref{fig:microMatParam_capPress} and \ref{fig:microMatParam_relPermeab}. 
In the following these are always denoted as splines or spline functions. 
The other two simulations use the curves derived from the Van Genuchten functions depicted in Figure \ref{fig:microMatParam_capPress} and \ref{fig:microMatParam_relPermeab}. 
For both types of pressure functions, we used either a constant contact angle of \SI{18}{\degree} or the dynamic contact angles from the fit in Fig.~\ref{fig:dyn_CA}. 

In the pore-morphology method used in the calculation of the effective material parameters, the contact angle only influences the capillary pressure function as a multiplicative factor $\text{cos} (\theta)$. Consequently, the effective material parameters can be transformed from an old contact angle $\theta^{\text{old}}$ to a new contact angle $\theta^{\text{new}}$ by simply multiplying the capillary pressure function of the old contact angle with the fraction $\frac{\text{cos} (\theta^{\text{new}})}{\text{cos} (\theta^{\text{old}})}$. This is used in the simulations with dynamic contact angles. In that case $\theta^{\text{old}} = 18^{\circ}$ and $\theta^{\text{new}}$ is calculated with the local velocity magnitude and the functional dependence of the contact angle to the velocity depicted in Figure \ref{fig:dyn_CA}. 

\begin{figure}
	\centering
	\includegraphics[width=0.15\linewidth]{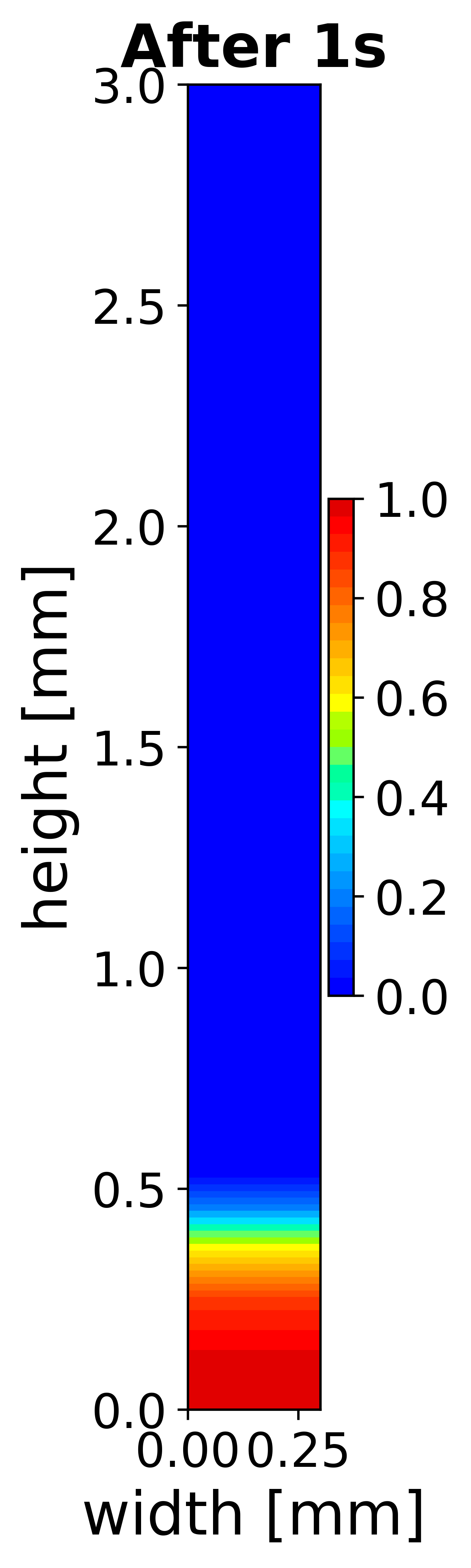}
	\includegraphics[width=0.15\linewidth]{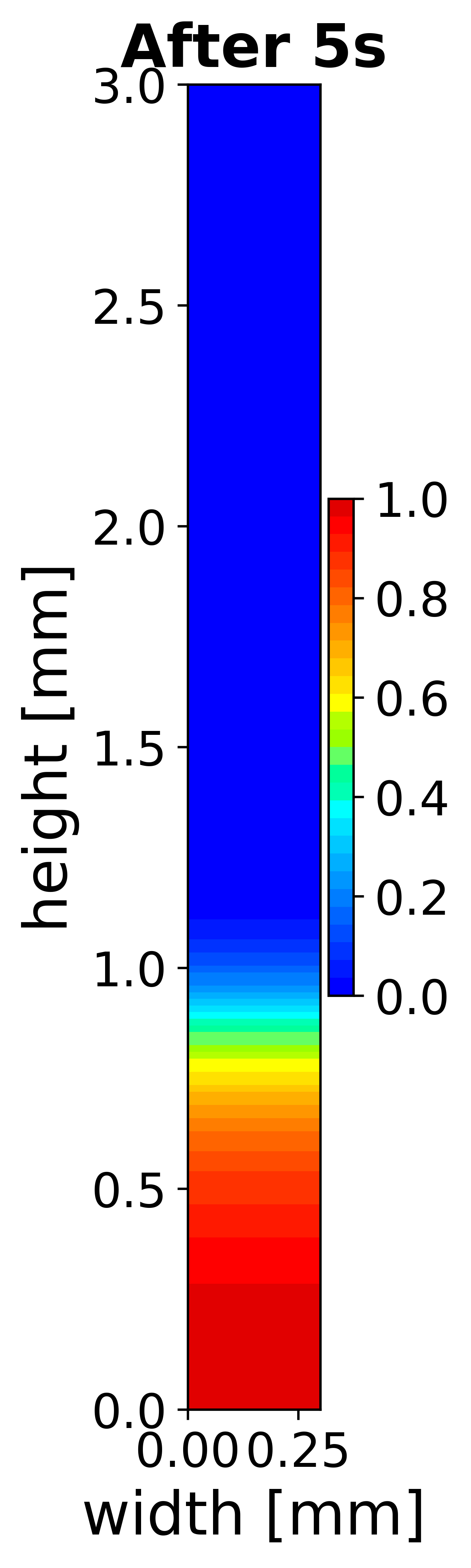}
	\includegraphics[width=0.15\linewidth]{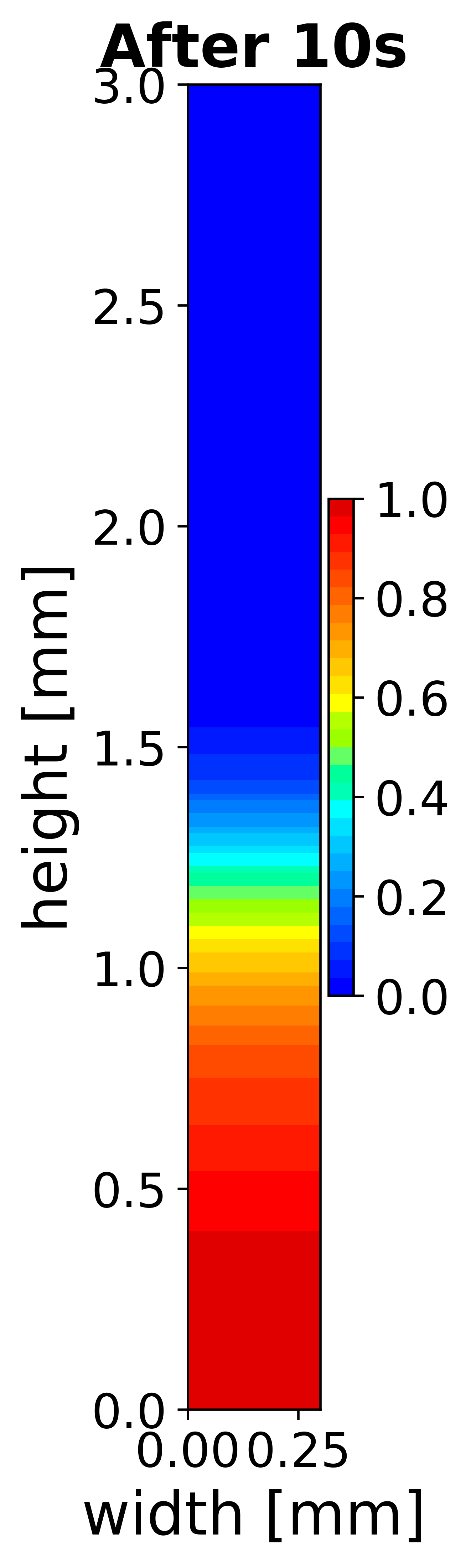}
	\includegraphics[width=0.15\linewidth]{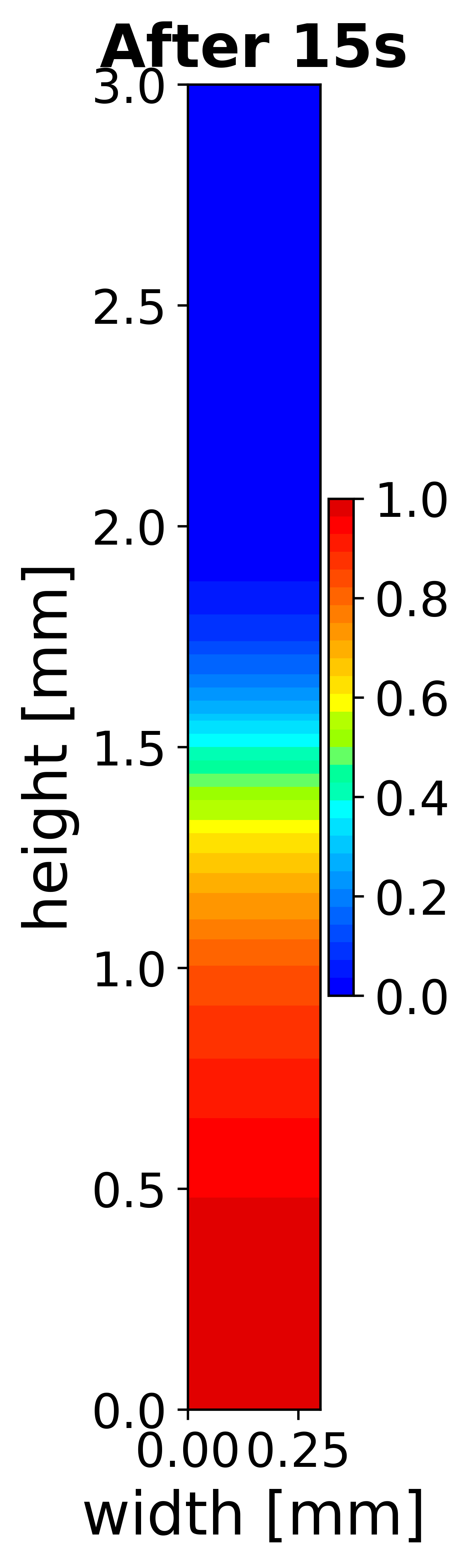}
	\caption{\label{fig:simResult2DColorplot} Presented is the saturation field at various times of the simulation using the static contact angle and the spline functions. The flow front is diffusive as it is expected for a capillary pressure driven flow. }
\end{figure}

\begin{figure}
	\centering
	\includegraphics[width=1\linewidth]{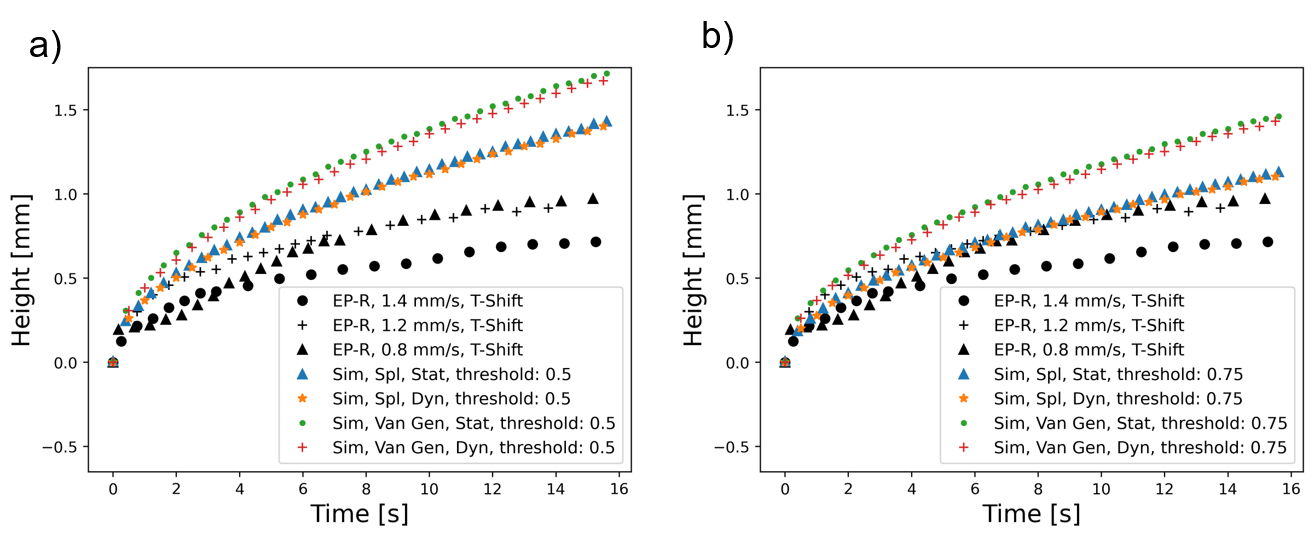}
	\caption{\label{fig:compSimToExp_flowFrontBound075} In these images the experimental data of Figure \ref{fig:dyn_CA} are compared to the simulation results. The experimental data are time-shifted to a flow front height of $0$ at time $0$. The heights of the flow front of the simulation results are computed using a threshold saturation of a) $0.5$ and a threshold saturation of b) $0.75$. The images show the simulation results for static and dynamic contact angles and for spline and Van Genuchten effective parameters.  }
\end{figure}

In Figure \ref{fig:simResult2DColorplot} the saturation field at four different times of the simulation using the static contact angle and the spline functions is presented.
To compare the simulation results to the experimental results, shown in the upper image of Figure \ref{fig:CapillaryRise}, the height of the flow front at each timestep is needed. In the simulation results there is no clear flow front because the saturation field is strongly diffusive. 
To get a value of the height of the flow front at each time we define a threshold saturation $S_w^{Tr} \in [0,1]$. The height of the flow front is then calculated by finding the largest height at which the saturation is still equal or higher than $S_w^{Tr}$. 
By this procedure, the data points in Figure \ref{fig:compSimToExp_flowFrontBound075} are computed. The experimental measurements shown in Figure \ref{fig:CapillaryRise} do not start at the height $0$, but the simulation results do. To address this inconsistency, we apply a time-shift to the experimental data. Therefore, we fit a square root function with a time shift, of the form
\begin{flalign} \label{eq:darcySol_ansatzFctTimeShift}
	a \sqrt{t + t_{shift}} ,
\end{flalign} 
to the first few data points of each experiment, inspired by the Lucas-Washburn solution \cite{Lucas1918, PhysRev.17.273}. 
That means the scalar parameters $a > 0$ and $t_{shift} > 0$ are fitted for the three experiments. The time values of the experimental data are then shifted by their respective value of $t_{shift}$. For the first few time steps, gravity plays a minor role, as capillary forces dominated. 
Under these conditions, the Lucas-Washburn equation approximates the height of the flow front as a square root function of time $t$. This is the reason why we use the function of Equation \eqref{eq:darcySol_ansatzFctTimeShift} to fit the first few data points of the measurements. 
The resulting comparison of simulations and experiments is shown in Figure \ref{fig:compSimToExp_flowFrontBound075} for threshold saturations $0.5$ and $0.75$.

The influence of the dynamic contact angle on the results is low, which is due to the fact that the rise velocity is most of the time in the lower range of the velocities tested in Fig.~\ref{fig:dyn_CA}. 
The direct use of the microstructure results in capillary pressure and relative permeability provides a good agreement with the measurements, which also show a certain range of variation.
The fit to the Van Genuchten curve, on the other hand, deviates more significantly, but also shows the correct qualitative behavior.  
Of course, the choice of threshold for saturation also has an influence on the exact course of the saturation curve, as the simulation produces a diffuse saturation front.
Overall, there is good agreement between the measurement and simulation results.

\section{Conclusion}

In this study, we presented a combination of experimental and simulated investigations into the wetting dynamics of a carbon fibre roving in an epoxy resin system, using a custom optical wetting setup. Microscale experiments were conducted to characterise the individual materials and determine the key parameters required for the simulation approach. At the mesoscale, the dynamic rise of the resin flow front inside the roving was observed optically, and changes in the roving's geometry were analysed post-curing. The time-dependent infiltration of the resin into the roving could be modelled quantitatively.

We found that the contact angles between the resin and the reinforcing fibres depend on the infiltration speed inside the roving and are not constant. However, the dynamic contact angle was found to have little influence on the capillary rise simulations considered. Instead, the exact shape of the capillary pressure function and the relative permeabilities had a significant impact on the simulation results. These parameters, in turn, depend on the microscale geometry of the roving, which was shown to depend on the sizing chemistry and distribution, surface roughness, and the number and shape of filaments in the roving. This geometry and its porosity were found to change during the infiltration process. 

It was possible to bring experimental data and simulation results into good agreement. However, the simulations define the flow front by selecting a specific saturation within a diffusive contact line, whereas the experiments involved a defined flow front. More time needs to be invested in resolving the complex contact line behaviour inside the roving material. Nevertheless, these trials present an approach that investigates complex wetting phenomena in composite manufacturing processes that are close to real-world scenarios, thus providing an extended insight into the processing behaviour of a widely used and relevant class of materials. Optimising processing parameters, as well as simulating composite materials and their manufacture, as demonstrated in this study, will remain important in the coming years. 

\pagebreak

\appendix
\section{Appendix}
\label{app1}

\begin{figure} [h]
	\centering
	\includegraphics[width=0.75\linewidth]{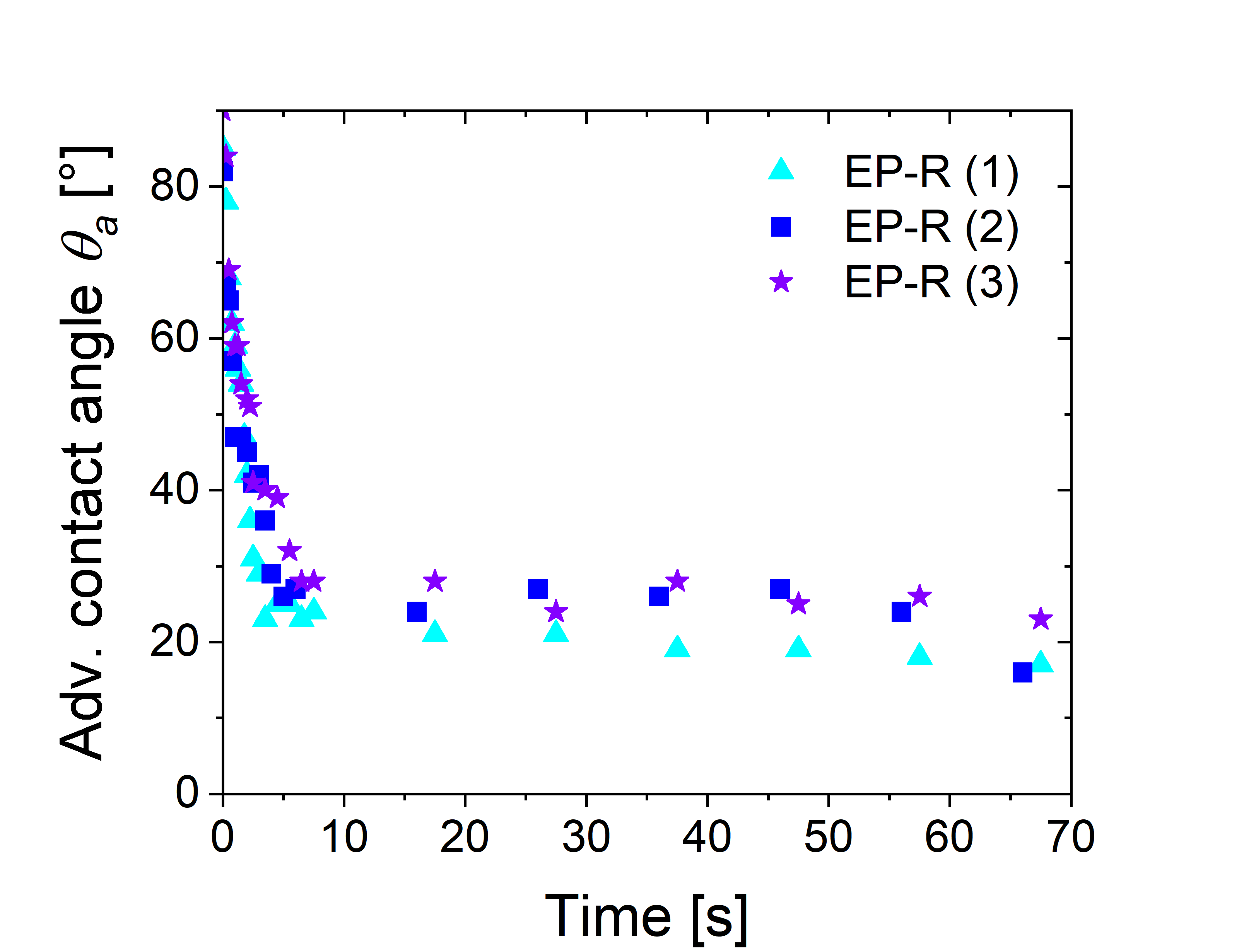}
	\caption{\label{fig:StaticContactAngleRelaxation}
		Following the sudden cessation of motion of the dip coater, the contact angle initially exhibited fast relaxation, followed by slow relaxation. We determined the contact angle to be \SI{60}{\second} after the end of the fast relaxation. This process is shown for three different samples of the same roving (EP-R).  The roving was driven into the resin at high velocity such that the dynamic advancing contact angle was above \SI{90}{\degree}. The time axis starts when the contact angle drops below \SI{90}{\degree} and becomes observable in our setup. The averaged final values are given in Table \ref{tab:AdvStaticContacAngles}}.
\end{figure}

\begin{figure*} [h]
	\centering
	\includegraphics[width=0.75\linewidth]{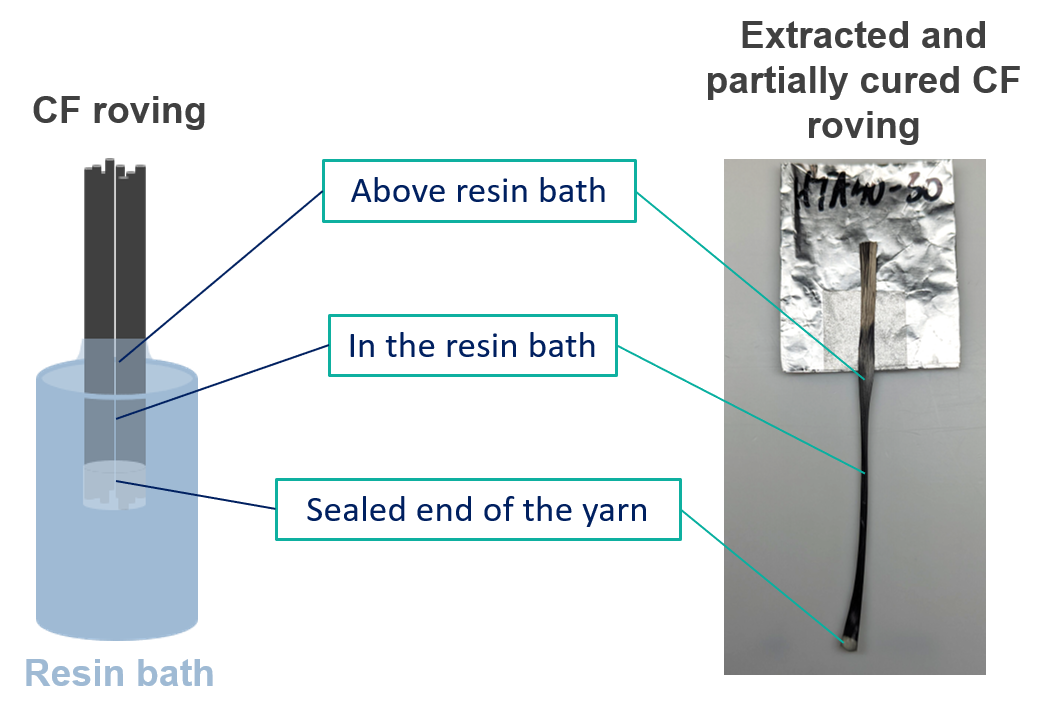}
	\caption{\label{fig:cross-sections} The three positions at which a cross section of the extracted and partially cured carbon fiber roving was produced: at the point right above the resin bath, inside the resin bath and inside the sealed end of the roving.}
\end{figure*}







\pagebreak

\end{document}